\documentclass[journal]{IEEEtran}
\usepackage{graphicx}
\usepackage{multirow}
\usepackage{subfig}
\usepackage{array}
\graphicspath{{Figures/}}
\usepackage{amsmath}
\usepackage{amssymb}
\usepackage{amsfonts}
\usepackage{epsfig}
\usepackage{float}
\usepackage{bm}
\usepackage{algorithm}
\usepackage{algorithmicx}
\usepackage{algpseudocode}
\usepackage{booktabs}
\usepackage{subfig}
\usepackage{caption}
\usepackage{rotating}
\usepackage{longtable,caption,float}
\usepackage{pifont}
\usepackage{color}
\ifCLASSINFOpdf
\else
\fi

\begin{document}
	\title{HPRN: Holistic Prior-embedded Relation Network for Spectral Super-Resolution}
	
	\author{Chaoxiong~Wu,~\IEEEmembership{Student Member,~IEEE,}
		Jiaojiao~Li,~\IEEEmembership{Member,~IEEE,}
		Rui~Song,~\IEEEmembership{Member,~IEEE,}
		Yunsong Li~\IEEEmembership{Member,~IEEE,}
		Qian Du~\IEEEmembership{Fellow,~IEEE}
		\thanks{This work was supported in part by the National Nature Science Foundation of China (no. 61901343), the state Key Laboratory of Geo-Information Engineering, (no. SKLGIE2020-M-3-1), the China Postdoctoral Science Foundation (no. 2017M623124), the China Postdoctoral Science Special Foundation (no.2018T111019) and the Innovation Fund of Xidian University (no. 5001-20109215456). The project was also partially supported by the science and technology on space intelligent control laboratory ZDSYS-2019-03 and the Fundamental Research Funds for the Central Universities JB190107. It was also partially supported by the National Nature Science Foundation of China (no. 61571345, 61671383, 91538101, 61501346 and 61502367), the 111 project (B08038), and the Innovation Fund of Xidian University (no.10221150004). (Corresponding author: Jiaojiao Li)}
		\thanks{C. Wu, J. Li, R. Song and Y. Li are with the State Key Laboratory of Integrated Service Networks, Xidian University, Xi’an 710071, China; J. Li is also with CAS Key Laboratory of Spectral Imaging Technology, Xi’an 710119. (e-mail: jjli@xidian.edu.cn; cxwu1120@163.com; rsong@xidian.edu.cn; Ysli@mail.xidian.edu.cn). Q. Du is with the Department of Electronic and Computer Engineering, Mississippi State University, Starkville, MS 39762 USA (e-mail: du@ece.msstate.edu).}
		
		\thanks{Manuscript received April 19, 2005; revised August 26, 2015.}}
	
	\markboth{Journal of \LaTeX\ Class Files,~Vol.~14, No.~8, August~2015}%
	{Shell \MakeLowercase{\textit{et al.}}: Bare Demo of IEEEtran.cls for IEEE Journals}

	\maketitle
	
\begin{abstract}
Spectral super-resolution (SSR) refers to the hyperspectral image (HSI) recovery from an RGB counterpart. Due to the one-to-many nature of the SSR problem, a single RGB image can be reprojected to many HSIs. The key to tackle this ill-posed problem is to plug into multi-source prior information such as the natural spatial context-prior of RGB images, deep feature-prior or inherent statistical-prior of HSIs, etc., so as to effectively alleviate the degree of ill-posedness. However, most current approaches only consider the general and limited priors in their customized convolutional neural networks (CNNs), which leads to the inability to guarantee the confidence and fidelity of reconstructed spectra. In this paper, we propose a novel holistic prior-embedded relation network (HPRN) to integrate comprehensive priors to regularize and optimize the solution space of SSR. Basically, the core framework is delicately assembled by several multi-residual relation blocks (MRBs) that fully facilitate the transmission and utilization of the low-frequency content prior of RGBs. Innovatively, the semantic prior of RGB inputs is introduced to mark category attributes, and a semantic-driven spatial relation module (SSRM) is invented to perform the feature aggregation of clustered similar range for refining recovered characteristics. Additionally, we develop a transformer-based channel relation module (TCRM), which breaks the habit of employing scalars as the descriptors of channel-wise relations in the previous deep feature-prior, and replaces them with certain vectors to make the mapping function more robust and smoother. In order to maintain the mathematical correlation and spectral consistency between hyperspectral bands, the second-order prior constraints (SOPC) are incorporated into the loss function to guide the HSI reconstruction. Finally, extensive experimental results on four benchmarks demonstrate that our HPRN can reach the state-of-the-art performance for SSR quantitatively and qualitatively. Further, the effectiveness and usefulness of the reconstructed spectra are verified by the classification results on the remote sensing dataset. Codes are available at {\color{magenta}https://github.com/Deep-imagelab/HPRN}.
\end{abstract}

	
	\IEEEpeerreviewmaketitle
	\section{Introduction}
\IEEEPARstart
Hyperspectral imaging can cover a wider range of the electromagnetic spectrum than the ordinary RGB cameras \cite{harsanyi1994hyperspectral}. The captured HSIs usually contain a great quantity of spectral bands. Such abundant spectral information reflects intrinsic properties of objects, and promotes the rapid applications of HSIs in remote sensing \cite{bioucas2013hyperspectral, melgani2004classification}, medical diagnosis \cite{lu2014medical, wei2019medical}, image classification \cite{li2018hyperspectral, li2018classification}, target detection \cite{akbari2010detection, shi2020hyperspectral} to name a few. However, the current hyperspectral imaging technology exists some practical bottlenecks from the hardware perspective, which are sensitive to the acquisition of high-quality HSIs. Typically, the inescapable loss of obtaining an HSI cube with higher spectral resolution is either the worse time-resolution, such as whisk-broom and push-broom scanning \cite{green1998imaging, james2007spectrograph}, or lower spatial resolution, such as integrated devices based on variable-filter design \cite{schechner2002generalized}. This trade-off lies in that these equipments must operate in a 1D or 2D scanning mode to acquire 3D data. At the expense of sacrificing moderate spatial or spectral resolution, recent advances in snapshot HSIs imaging can accomplish the hyperspectral video shooting, which can merely meet the coarse-grained application analysis \cite{wagadarikar2008single, wagadarikar2009video, wang2016adaptive, tanriverdi2019dual}. Also, these relevant imaging facilities inevitably need to bear expensive hardware overhead and burden.

\begin{figure*}[!tbp]
%	\vskip -0.05in
	\centering
	\scalebox{1}
	{
		\begin{tabular}{@{}c@{}c@{}c@{}c@{}c@{}c@{}c@{}c@{}c}
			&$\text{Galliani\cite{galliani2017learned}}$ &$\text{MSCNN\cite{yan2018accurate}}$&$\text{UNet\cite{stiebel2018reconstructing}}$&$\text{HSCNN+\cite{shi2018hscnn+}}$&$\text{FMNet\cite{zhang2020pixel}}$& $\text{HRNet\cite{zhao2020hierarchical}}$ &$\text{Ours}$ \\
			\rotatebox{90}{\ \ \ \ \ \
				$\text{Spectra}$}\ \ 
			&{\includegraphics[width=2.42cm,height=2.42cm]{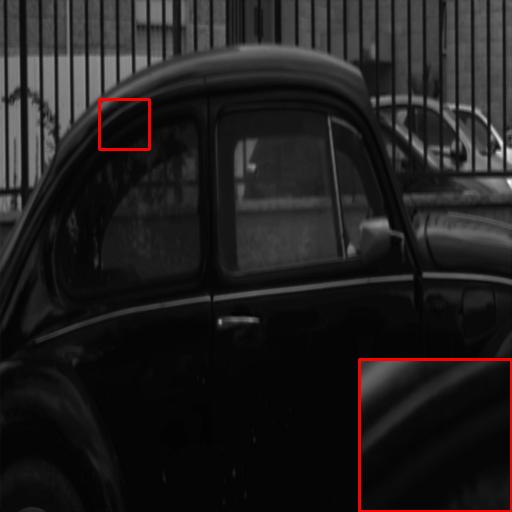}} \
			&{\includegraphics[width=2.42cm,height=2.42cm]{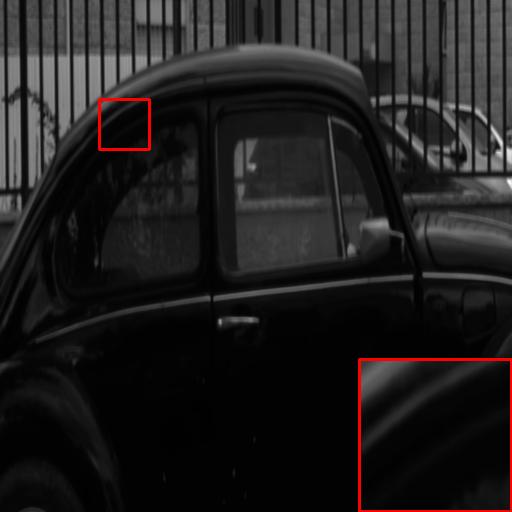}} \
			&{\includegraphics[width=2.42cm,height=2.42cm]{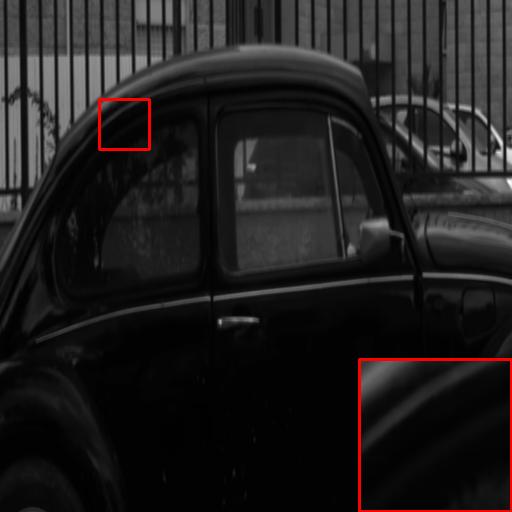}} \
			&{\includegraphics[width=2.42cm,height=2.42cm]{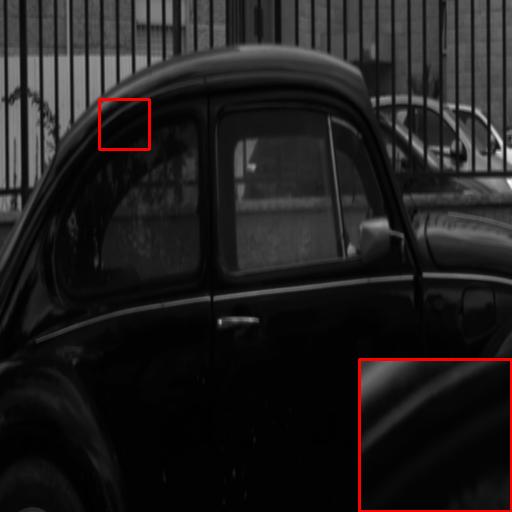}} \
			&{\includegraphics[width=2.42cm,height=2.42cm]{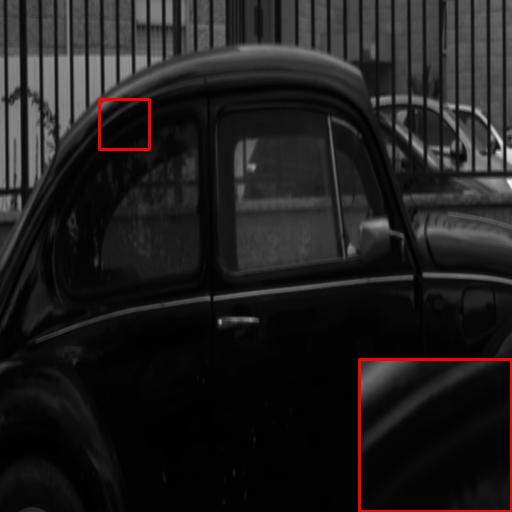}} \
			&{\includegraphics[width=2.42cm,height=2.42cm]{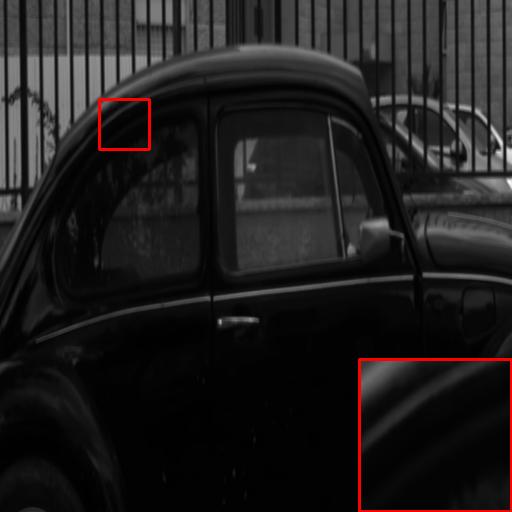}} \
			&{\includegraphics[width=2.42cm,height=2.42cm]{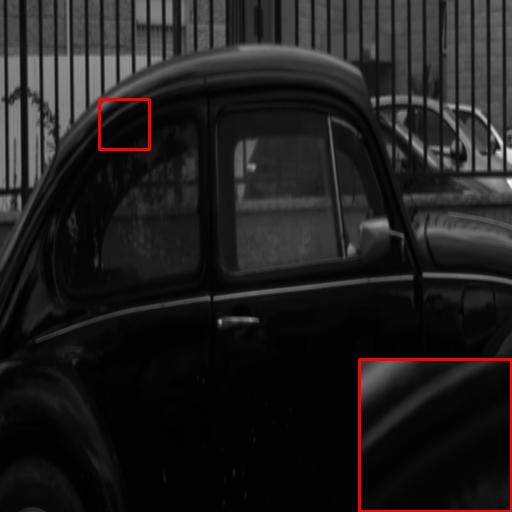}} \\
			\rotatebox{90}{\ \ \ \ \ \ \ 
				$\text{Error}$}\ \ 
			&{\includegraphics[width=2.42cm,height=2.42cm]{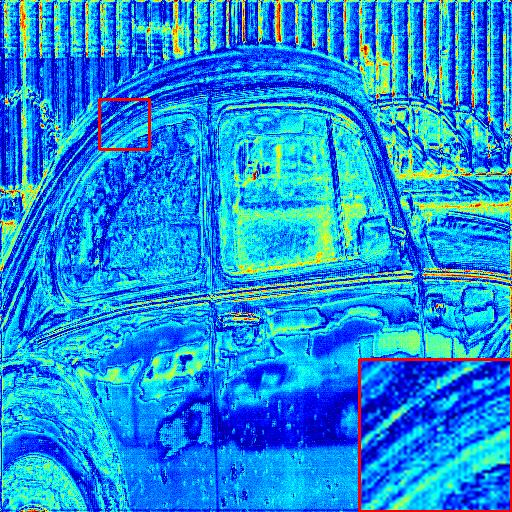}} \
			&{\includegraphics[width=2.42cm,height=2.42cm]{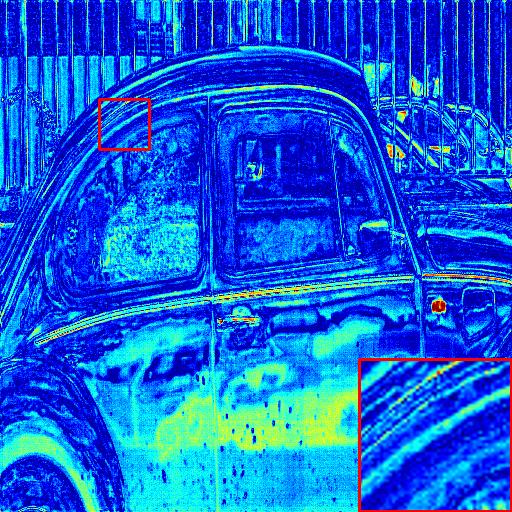}} \
			&{\includegraphics[width=2.42cm,height=2.42cm]{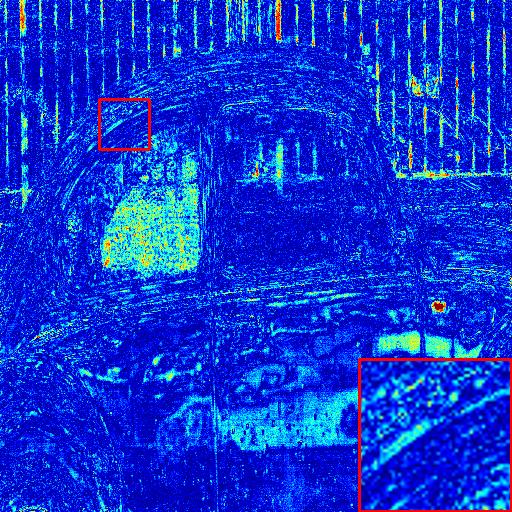}} \
			&{\includegraphics[width=2.42cm,height=2.42cm]{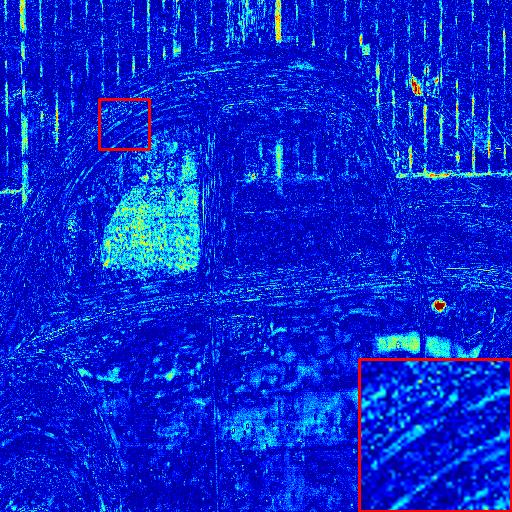}} \
			&{\includegraphics[width=2.42cm,height=2.42cm]{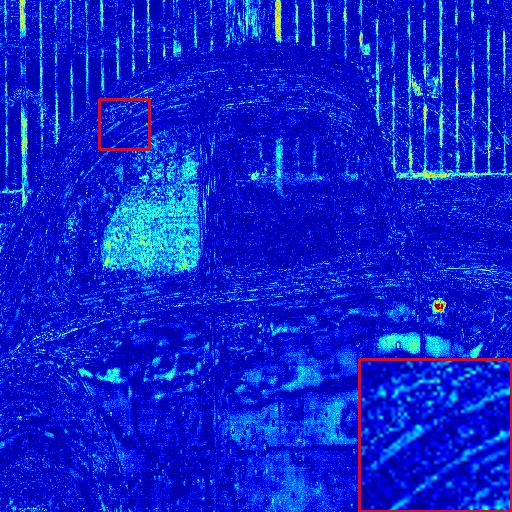}} \
			&{\includegraphics[width=2.42cm,height=2.42cm]{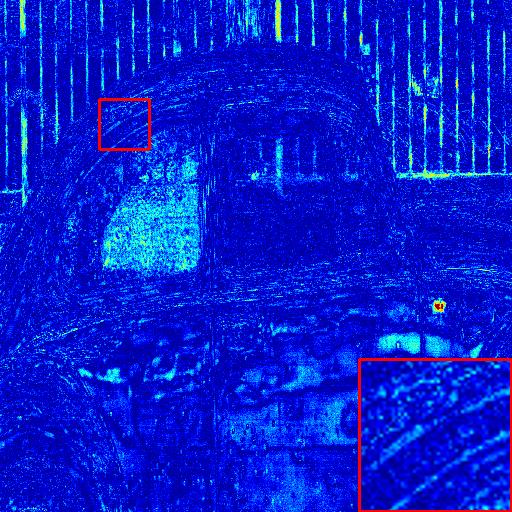}} \
			&{\includegraphics[width=2.42cm,height=2.42cm]{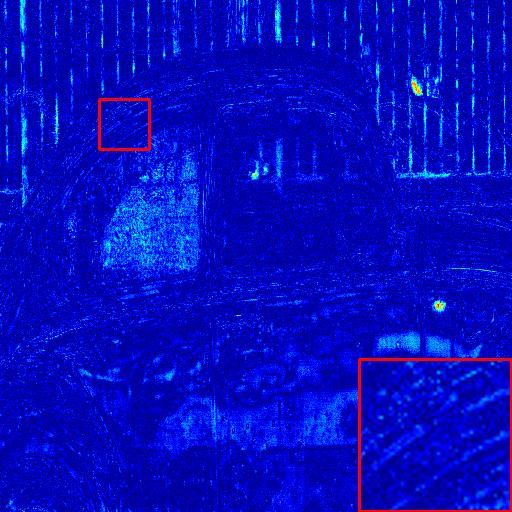}} \
			&{\includegraphics[width=0.2cm,height=2.42cm]{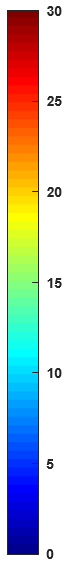}}\\
		\end{tabular} 
	}
	\caption{Visual comparison of the 520nm band on ``BGU\_HS\_00259'' HSI from the NTIRE2018 ``Clean'' dataset. The best view on the screen. 
	}
	\label{figure_intro}	
%	\vskip 0.05in
\end{figure*}

To cope with the above shortcomings of hardware factors, researchers turn their energies to the software perspective to excavate solutions, for example, computed tomography imaging spectrometers \cite{okamoto1991simultaneous}, compressed sensing techniques \cite{gehm2007single} and multiplexed illuminations \cite{park2007multispectral, parmar2008spatio, goel2015hypercam, oh2016yourself}, etc. However, these solutions sometimes rely on special environments, and require significant post-processing calculations to restore the complete spectral signatures. Later, some sparse coding schemes are proposed to directly utilize a single RGB image to recover the corresponding HSI \cite{robles2015single, arad2016sparse, aeschbacher2017defense}. Through leveraging statistical sparse priors of HSIs, they firstly build an over-complete dictionary, and then treat spectral reconstruction as a linear transformation from RGB images to HSIs. Advantageously, these algorithms not only create a convenient way to obtain HSIs, but also substantially reduces hardware costs from HSI imaging devices to RGB cameras. Unfortunately, it is an extremely ill-posed conversion to predict high-dimensional spectral stimuli from three-dimensional RGB signals. Thus, only involving the simple and hand-crafted sparse prior in the RGB-to-HSI transition, such sparse coding ways can not guarantee the authenticity and feasibility of the generated spectrum. More importantly, the linear transformation is not enough to characterize the hypothesis space of the underdetermined SSR issue.

In recent years, deep CNNs have been widely exploited in the SSR task \cite{galliani2017learned, rangnekar2017aerial, yan2018accurate, xiong2017hscnn, mei2020spatial}. To promote the development of SSR technology, New Trends in Image Restoration and Enhancement (NTIRE) has organized two SSR competitions in 2018 \cite{arad2018ntire} and 2020 \cite{arad2020ntire}, where numerous CNN-based methods have been proposed for hyperspectral recovery \cite{shi2018hscnn+, stiebel2018reconstructing, li2020adaptive, zhao2020hierarchical, peng2020residual, zhang2020pixel, li2021progressive}. Through extracting the abstract deep feature-prior from the dataset of RGB-HSI pairs, these approaches learn end-to-end nonlinear mappings between the RGB inputs and the HSI counterparts. Compared with the previous liner transition of sparse dictionary learning, CNN-based algorithms can greatly intensify the precision of estimated spectral signatures. Still, several drawbacks exist in the current CNN-based models. Especially, most of the CNN techniques almost only explore the general contextual prior of RGB inputs and learn the implicit deep-feature prior via a customized network. Due to considering the generic and limited priors, these networks can merely alleviate the redundancy of this ill-posed problem to a certain extent, and provide an acceptable but relatively low-precision recovered spectral from many alternative results. On this basis, there are also a few models to introduce more advanced deep feature-prior into SSR, like certain attention modules \cite{zhao2020hierarchical, nathan2020light} to improve the learning ability of CNNs and acquire a smoother solution domain from the RGB input to HSI output. However, existing attention-based modules usually adopted global-averaging scalars as channel-wise squeezers, which probably failed to characterize the entire channel information, since global-averaging was easily distracted by background clutter. And these approaches rarely further combine superior knowledge, such as the semantic prior of RGB inputs and intrinsic statistical prior of HSIs, making SSR performance still limited. Besides, there are some SSR researches to integrate the known camera spectral sensitivity prior \cite{li2021deep, stiebel2020enhancing}, but this kind of prior is not always available in reality.
	
Towards the aforementioned problems, we take full advantage of multi-source prior information for SSR and propose a novel holistic prior-embedded relation network (HPRN). To be specific, the backbone is repeatedly stacked via several multi-residual relation blocks (MRBs). Profiting from such multi-residual connection paradigm, the low-frequency context prior of RGB images is adequately utilized through the deep network. Based on the fact that local category attributes and scene distributions are consistent between the given RGB images and associated HSIs spatially, a semantic-driven spatial relation module (SSRM) is well-designed innovatively at the tail of the presented HPRN. Take embedded semantic prior of RGB inputs as the category indexes for resolving HSIs, this module can carry out the feature aggregation across the similar spectral signatures using a semantic-based relation matrix. Such SSRM can selectively model the correlation learning of category-consistent patterns while reducing the attention of discrepant ones, which further effectively refines the quality of coarse reconstructed spectra. In addition, one local-range-averaging multiple-number vector can provide a more powerful expression than one global-averaging single-number scalar in terms of representing the channel of a deep feature. In this respect, we investigate a transformer-based channel relation module (TCRM), which leverages certain vectors instead of scalars as the squeezers of channel-wise relation to extract feature interrelations. Coupled with Transformer-style feature interactions, our TCRM can make the mapping function more robust, potentially regularizing and smoothing the one-to-many solution space. Attentively, band-wise correlations of HSIs are intrinsic and non-ignorable priors determined by the hyperspectral imaging principle. To reinforce hyperspectral band-wise statistical correlations and spectral continuity, the second-order prior constraint (SOPC) is added to the loss function as an auxiliary term to assist the network training process. Experimental results on four prevalent SSR benchmarks demonstrate that our HPRN can achieve the state-of-the-art performance quantitatively and qualitatively. Meanwhile, the classification results on the remote sensing dataset validate the effectiveness and usefulness of the reconstructed spectrum. As shown in Fig. \ref{figure_intro}, our HPRN can acquire better visual quality and less reconstruction errors over other superior SSR methods.

The main contributions of this paper can be summarized as follows:
	
\begin{itemize}
\item We present a novel holistic prior-embedded relation network (HPRN) for SSR. Multi-source and abundant priors including spatial contexts of RGB images, semantic information of RGB signals, deep feature-prior and band-wise correlations of HSIs are incorporated into the end-to-end mapping function, which can effectively alleviate the ill-posedness of the underconstrained SSR problem, further promoting the accuracy and fidelity of recovered HSIs.

\item Through embedding semantic prior of RGB inputs, a trainable semantic-driven spatial relation module (SSRM) is well-designed innovatively to perform the feature aggregation among the clustered similar characteristics of the reconstructed HSIs. Such SSRM can accomplish semantic-guided correlation learning of category-consistent patterns, and effectively achieve spectral optimization of the coarse estimation at the end of our HPRN.

\item We develop a transformer-based channel relation module (TCRM), which leverages certain vectors rather than scalars as the descriptors of channel-wise relations to explore feature interdependences. Together with Transformer-style feature interactions, this module can obtain more discriminative learning power to produce a more robust and smoother one-to-many mapping function potentially. 

\item The second-order prior constraint (SOPC) is additionally incorporated into the loss function to assist the network learning, playing a role in maintaining the hyperspectral band-wise statistical correlations and spectral continuity. Mathematically, the SOPC term can further assist the L1 loss to make the space of possible one-to-many mappings smaller to implement a high-precision spectral recovery.

\item Extensive experimental results demonstrate that our HPRN can surpasses the state-of-the-art SSR methods under multiple evaluation metrics on four established benchmarks. Also, the effectiveness and usefulness of the reconstructed spectrum are proved by the classification results on the remote sensing dataset.
\end{itemize}

\section{Related Works}
\subsection{Traditional Methods}
In the last few decades, hyperspectral imaging has been successfully proven to be beneficial for many applications in the fields of environment, remote sensing, geography and so on. With the progress of science and technology, hyperspectral imaging has also been continuously developed and upgraded. Traditional whisk-broom or push-broom imagers employ the point-by-point or line-by-line scanning manners to collect the scene radiations and reflections \cite{green1998imaging, james2007spectrograph}. Although these devices can acquire contiguous spectral bands among the larger-range electromagnetic spectrums, they are extremely slow and time-consuming. To fulfill fast acquisition of HSIs, coded aperture snapshot hyperspectral cameras are devised to perform spectral imaging of a dynamic scene at a video rate \cite{wagadarikar2008single, wagadarikar2009video, wang2016adaptive, tanriverdi2019dual}. Despite the faster speed, there is usually a loss of spatial or spectral resolution. Unavoidably, the hardware components of the imaging equipment are still expensive. Thus conventional RGB cameras are tried to capture the spectral signatures of natural objects \cite{park2007multispectral, parmar2008spatio, goel2015hypercam, oh2016yourself}. For example, Goel \emph{et al.} \cite{goel2015hypercam} designed a low-cost multispectral system with a digital camera and a software approach that automatically analyzes the scene under controlled lighting. Oh \emph{et al.} \cite{oh2016yourself} built a framework for reconstructing HSIs by using multiple RGB cameras, and introduced an algorithm to combine and convert these different RGB measurements into a single HSI. However, such methods have to face with a huge computational burden.

Afterwards, some researchers proposed sparse coding approaches for reconstructing the corresponding HSI from a single RGB image. Firstly, Robles-Kelly \cite{robles2015single} extracted a set of prototypes from the training set and utilized sparse coding to predict the irradiances of HSIs. Then Arad and Ben-Shahar \cite{arad2016sparse} published a large database of natural HSIs for SSR, and learned an over-complete dictionary via the K-SVD algorithm \cite{aharon2006k}. Given an RGB pixel, the dictionary representation of hyperspectral signatures would be computed with the orthogonal match pursuit iterations \cite{pati1993orthogonal}. Purposefully, Aeschbacher \emph{et al.} \cite{aeschbacher2017defense} reimplemented and pushed the performance of the work \cite{arad2016sparse}, then introduced a novel shallow network based on the super-resolution method  \cite{timofte2014a+}. Eventually, the time-consuming and laborious hyperspectral imaging is transformed into the spectral reconstruction from an RGB signal cost-effectively and quickly. Regrettably, the SSR problem is ill-posed, which means that multiple HSIs solutions can project the same RGB input. Since the above-mentioned methods only involve the hand-crafted sparse prior of HSIs, they have a hard time choosing a high-confidence and high-quality spectrum from many candidates. To acquire more accurate spectral estimation, the more priors should be plugged into the SSR process to relieve the underdetermination. 

\begin{figure*}
	\centering
	\includegraphics[width=0.83\textwidth]{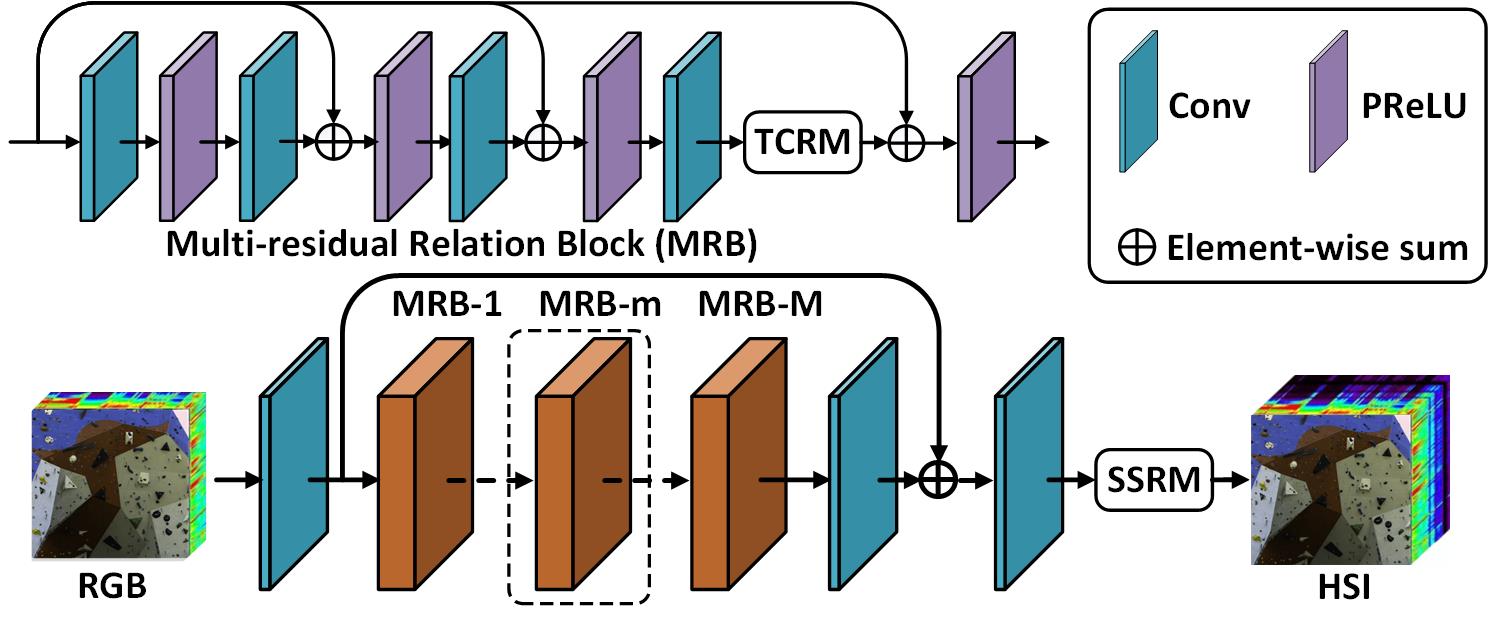}
	\caption{The network architecture of our presented holistic prior-embedded relation network (HPRN).}
	\label{figure_network}
\end{figure*}	
	
\subsection{CNN-based Models}
Deep CNNs have become the main-stream solution to many tasks, such as super-resolution \cite{niu2020single, mei2021image}, classification \cite{zhang2021information} and dehazing \cite{li2021multi, zhang2021semantic}, etc. In the early, Galliani \emph{et al.} \cite{galliani2017learned} proposed a pioneering CNN model for SSR. To boost the SSR accuracy, Yan \emph{et al.} \cite{yan2018accurate} presented a multi-scale CNN (MSCNN) to explicitly map the input RGB image to an HSI. Improved upon HSCNN \cite{xiong2017hscnn}, Shi \emph{et al.} \cite{shi2018hscnn+} built a hyperspectral reconstruction network HSCNN+, which achieved the first place on both the ``Clean'' and ``RealWorld'' tracks in the NTIRE 2018 Spectral Reconstruction Challenge \cite{arad2018ntire}. In 2020, NTIRE organized the second Spectral Reconstruction Challenge \cite{arad2020ntire}, and a new set of CNN-based models were proposed for SSR. Among them, Li \emph{et al.} \cite{li2020adaptive} developed an adaptive weighted attention network for more powerful feature expression. Zhao \emph{et al.} adopted the PixelShuffle layer as inter-level interaction, and proposed a 4-level Hierarchical Regression Network (HRNet) to explore context information for the spectral recovery. Moreover, Zhang \emph{et al.} \cite{zhang2020pixel} focused on the size-specific receptive field centered at each pixel, and investigated a pixel-aware deep function-mixture network (FMNet) to solve the SSR issue. Benefiting by the ability of CNN to implicitly extract abstract deep feature-prior from the dataset of large RGB-HSI pairs, CNN-based methods effectively regularized the one-to-many solution and heightened the accuracy of predicting spectral signatures.
	
To introduce more advanced deep feature-prior, there are also a few CNN-based models to design certain attention modules for learning interdependencies among intermediate features. The attention mechanism is generally regarded as an explanatory high-level prior, which imitates the human visual system that automatically captures salient parts of objects \cite{vaswani2017attention}. In the deep CNNs, the attention block can explore the correlation between deep features and adaptively focus on more informative ones. This can boost the learning power of the network, and make the end-to-end function smoother than plain CNNs. Nathan \emph{et al.} \cite{nathan2020light} presented a light weight residual dense attention net based on attention mechanisms. Li \emph{et al.} \cite{li2020hybrid} put forward the 2D channel and 3D band-wise attention modules in their work. Peng \emph{et al.} \cite{peng2020residual} devised a pixel attention module to rescale pixel-wise features in all feature maps. These attention-based blocks were accustomed to adopt the scalars computed by the global average pooling as channel-wise squeezers. Nevertheless, this global-averaging scalar probably failed to characterize the entire channel information, since it was easily distracted by background clutter and anomaly. Thus, to acquire the stronger feature representations, we replace scalars with certain vectors to learn feature interdependences and develop a transformer-based channel relation module (TCRM). Meanwhile, based on that the Transformer model has recently proved to be an effective tool in the computer vision \cite{chen2021pre, wu2021cvt}, we introduce Transformer-style feature interactions into the proposed TCRM to enhance the expression ability. Further, the one-to-many mapping space is regularized indirectly, thus we can obtain a more robust RGB-to-HSI solution.

Moreover, several related works exploited prior category information of specific objects into the spectral reconstruction. Han \emph{et al.} \cite{han2018spectral} employed the unsupervised clustering to divide each RGB image into multiple classes, and established the nonlinear spectral mapping for each class. Yan \emph{et al.} \cite{yan2020reconstruction} labeled a C2H-Data laboriously and directly loaded the semantic prior information into the intermediate layers of their C2H-Net. However, the former inefficiently performs the classification and reconstruction tasks step by step, while the latter relies on labor-intensive manual labeling. Innovatively, this semantic prior from RGB signals is subtly plugged into our network, and a semantic-driven spatial relation module (SSRM) is meticulously designed at the end of the HPRN. Such SSRM can carry out correlation learning and feature aggregation across category-consistent ranges, where the category indexes are assigned by the built-in semantic prior knowledge of RGB signals. Without extensive manual labeling and inefficiently performing spectral mapping for each class, the SSRM module can fulfill the spectral optimization of the coarse estimation, which can jointly work with the HSIs recovery process. With additional statistical band-wise correlations prior of HSIs, our HPRN can availably mitigate uncertainty of the underdetermined SSR problem, thus increasing the precision of estimating spectra.

\begin{figure*}
	\centering
	\includegraphics[width=0.85\textwidth]{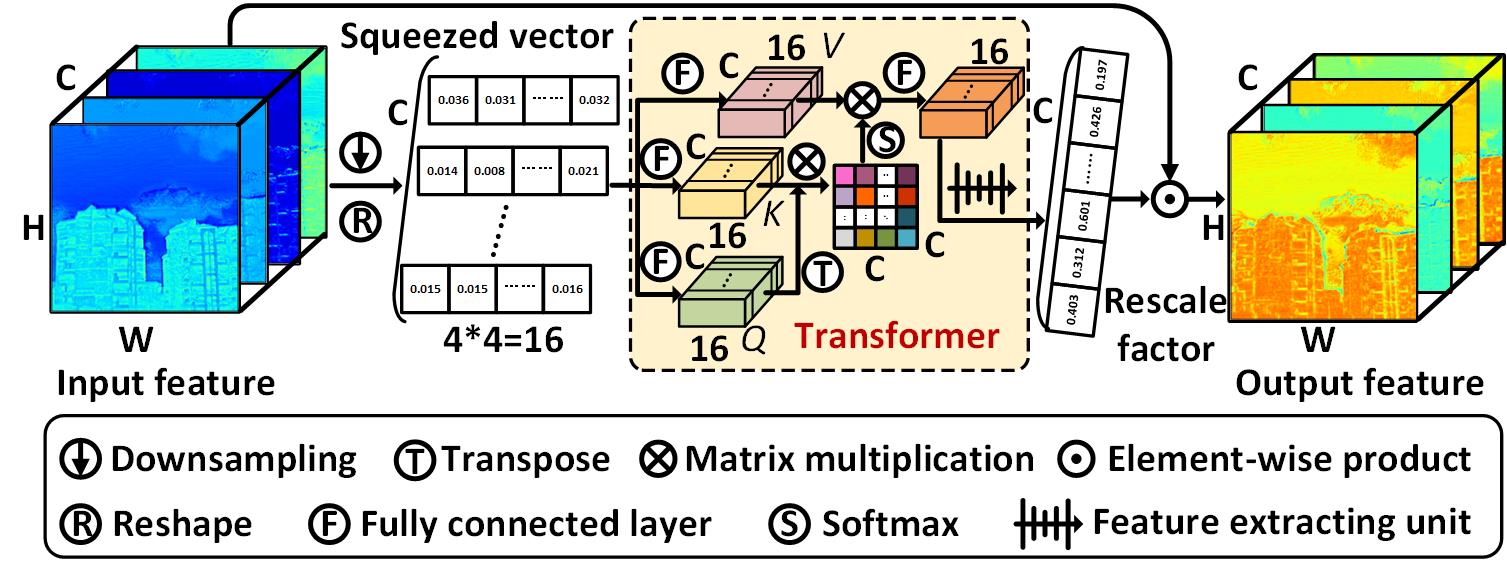}
	\caption{The diagram of the developed transformer-based channel relation module (TCRM).}
	\label{figure_tcrm}
\end{figure*}

\section{The Proposed Method}
\subsection{Preliminaries and Motivation}
Let ${\textbf{I}_{RGB}\in \mathbb{R}^{3\times H \times W}}$ and ${\textbf{I}_{HSI}\in \mathbb{R}^{B\times H \times W}}$ denote the given RGB image and corresponding ground HSI, where $B$, $H$ and $W$ are the band, height and width of an HSI cube. When the camera spectral sensitivity prior $\boldsymbol{\Phi}\in \mathbb{R}^{3\times B}$ is known, the RGB signal can be converted from the HSI measurement as follows \cite{fu2018joint}: 
\begin{equation}
\textbf{I}_{RGB}=\boldsymbol{\Phi} \textbf{I}_{HSI}.
\end{equation}
The conversion from $\textbf{I}_{HSI}$ to $\textbf{I}_{RGB}$ is a projection from higher dimension to lower one. Conversely, recovering HSIs from RGB inputs is obviously an ill-posed problem, indicating that one RGB image can be reprojected to multiple HSIs. Thus, integrating a large amount of prior information into SSR is beneficial to regularize the solution space, which can further advances the fidelity of reconstructed HSIs. Since the camera spectral sensitivity prior is sometimes unknown in reality, mathematically, the super-resolved HSI $\textbf{I}_{SSR}\in \mathbb{R}^{B \times H\times W}$ can be obtained by the following formula
\begin{equation}
\textbf{I}_{SSR}=\boldsymbol{\Psi}(\textbf{I}_{RGB}, \mathbb{P}(\cdot)),
\end{equation}
where $\mathbb{P}(\cdot)$ is the prior term containing various available prior information from the SSR task. In our proposed HPRN, we merge abundant priors including semantic categories of RGB signals, more advanced deep feature-prior and band-wise correlations of HSIs besides the general contextual prior of RGB inputs, which can constrain and optimize the one-to-many solution domain, further improving the quality of reprojected spectra.

\subsection{Network Architecture}
\label{Network Architecture}
Fig. \ref{figure_network} gives an illustration of the presented HPRN. Initially, a convolutional layer is placed for shallow feature-prior learning from the input RGB context-prior
\begin{equation}
\mathbf{F}_{0}=H_{SFP}\left(\mathbf{I}_{RGB}\right),
\end{equation}
where $H_{SPF}(\cdot)$ stands for the convolution function. Then the obtained shallow feature-prior $\mathbf{F}_{0}$ is fed into the backbone network for the deep feature-prior extraction. Concretely, the basic framework is composed of several MRBs. The whole process is expressed as
\begin{equation}
	\mathbf{F}_{DFP}=H_{MRBs}\left(\mathbf{F}_{0}\right),
\end{equation}
where $H_{MRBs}(\cdot)$ denotes the deep feature-prior extraction. Compared with the classical residual module, our MRB adopts the multi-residual connections, which can make the best of the low-frequency context-prior of RGB images. Next, the deep feature-prior $\mathbf{F}_{DFP}$ is injected into another convolutional layer, and interacts with the shallow feature-prior $H_{SFP}$ to form a global residual summation (GRS).
\begin{equation}
	\mathbf{F}_{GRS}=H_{GRS}\left(\mathbf{F}_{DFP}\right)+\mathbf{F}_{0},
\end{equation}
where $H_{GRS}(\cdot)$ is the convolutional weights. Such operation can avoid causing gradients to disappear or explode, and strengthen the stability of network training. At the end of our HPRN, we set up the reconstruction part consisting of a single convolution and the developed SSRM. The former adjusts the channel-wise number of the feature $\mathbf{F}_{GRS}$ equal to the ground ${\textbf{I}_{HSI}}$. And a coarse HSI recovery $\mathbf{F}_{Coa}$ can be acquired, which would be refined via the latter SSRM module:
\begin{equation}
	\mathbf{I}_{SSR}=H_{Rec}\left(\mathbf{F}_{GRS}\right)=H_{HPRN}(\mathbf{I}_{RGB}),
\end{equation}
where $\mathbf{I}_{SSR}$ and $H_{HPRN}(\cdot)$ denote the final reconstructed HSI and the function of our presented HPRN, respectively. The detailed implementation of SSRM is described in Section \ref{SSRM}.

\begin{figure*}
	\centering
	\includegraphics[width=0.83\textwidth]{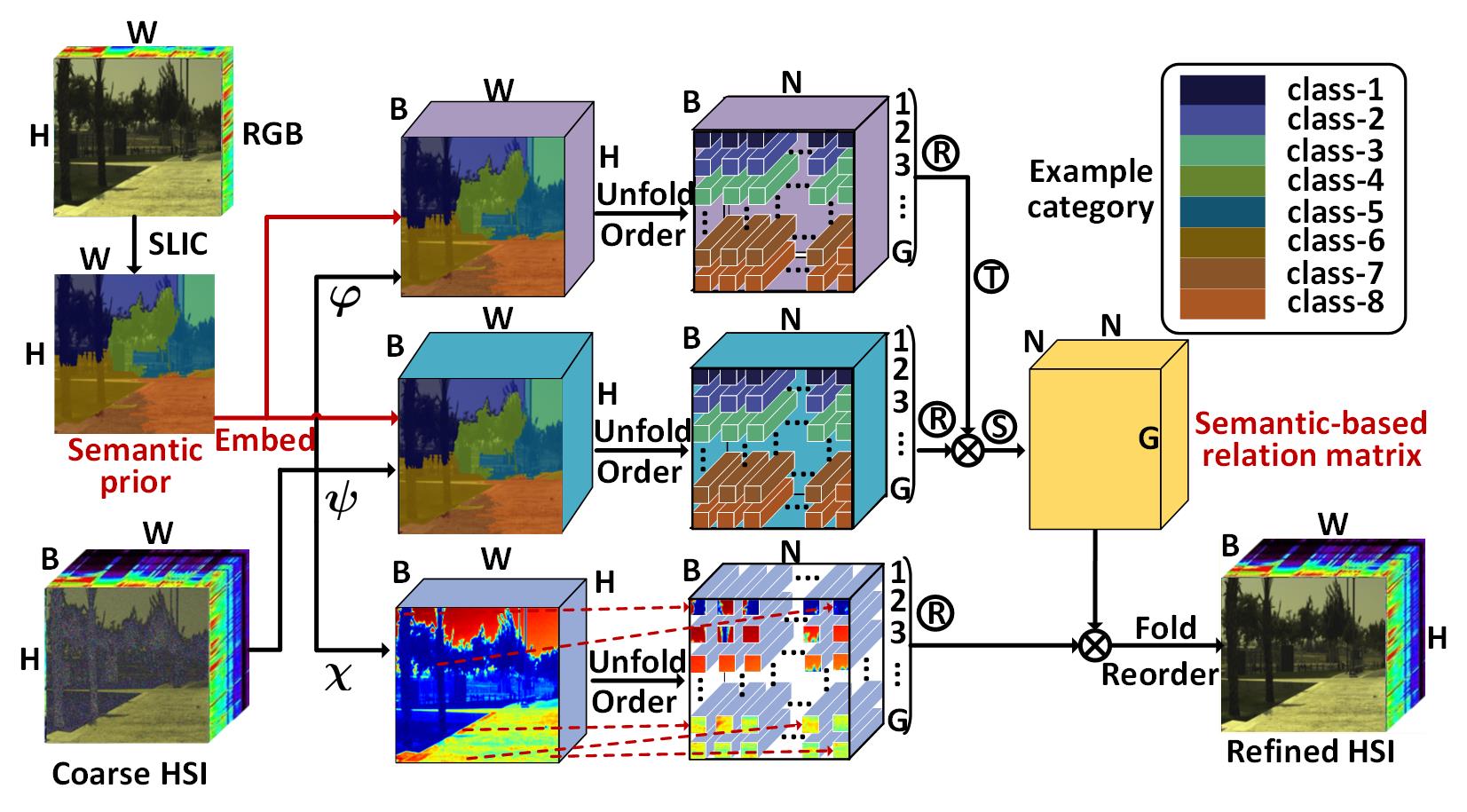}
	\caption{The diagram of the well-designed semantic-driven spatial relation module (SSRM).}
	\label{figure_ssrm}
\end{figure*}

\subsection{Transformer-based Channel Relation Module (TCRM)}
%Recently, there were a few SSR models involving certain attention blocks to explore feature dependencies across different channels \cite{zhao2020hierarchical, nathan2020light}. This attention mechanism can be interpreted as a more advanced deep feature-prior that can emphasize more informative features and enhance the learning power of the plain CNN. Popularly, existing attention-based SSR methods are always accustomed to adopting the global average pooling to obtain a single number as the channel-wise relational descriptor. However, this global-averaging scalar probably fails to characterize the entire channel information, since it is easily distracted by background clutter and anomaly. Therefore, our presented TCRM module employs a vector with a certain length as the channel representation, supporting it to be more discriminative.

In this part, we would introduce the developed TCRM module in detail. Ideally, given an input feature map $\mathbf{F}\in \mathbb{R}^{C\times H\times W}, \mathbf{F}=[\mathbf{f}_1, \mathbf{f}_2, \cdots, \mathbf{f}_C]$, where $C$ is the number of channels. As seen in Fig. \ref{figure_tcrm}, the feature $\mathbf{F}$ is firstly divided into $4\times4$ grids evenly along the spatial dimension. Then the local-range-averaging pooling is employed to generate the squeezed vectors $\mathbf{F}_{SV}\in \mathbb{R}^{C\times4\times4}$ for all channels.
\begin{equation}
\mathbf{F}_{SV}=H_{LAP}\left(\mathbf{F}_{in}\right),
\end{equation}
where $H_{LAP}(\cdot)$ denotes the local average pooling function. Compared with global-averaging single-number scalar to characterize the entire channel information, one local-range-averaging multiple-number vector can provide a stronger representation. The reason is that the prominent targets generally exist in a certain local position, and the global-average downsampling may be biased by the background noise and anomaly in other positions. Moreover, from the perspective of information theory, a vector of multiple numbers contains more information than a scalar of a single number. To capture non-local (also local) relations among channels, a transformer-based block is applied to perform the feature interaction. Concretely, we adopt its core component i.e., the multi-head self-attention layer, which can be invoked directly in the current Pytorch framework \cite{paszke2017automatic}. Formally, the squeezed vectors $\mathbf{F}_{SV}\in \mathbb{R}^{C\times4\times4}$ are reshaped into $\mathbf{F}_{SV}^{'}\in \mathbb{R}^{C\times(4\times4)}$, which are subsequently fed into the transformer-based block:
\begin{equation}
\mathbf{F}_{Tr}=H_{MSA}\left(\mathbf{F}_{SV}^{'}\right)=\operatorname{softmax}\left({Q\otimes K^{T}}\right)\otimes V,
\end{equation}
where $H_{MSA}(\cdot)$ denotes the multi-head self-attention module including four fully connected layers to produce query ($Q$), key ($K$), value ($V$) and output vectors. To obtain the rescaling factor for each channel, the output vectors $\mathbf{F}_{Tr}\in \mathbb{R}^{C\times(4\times4)}$ are firstly meaning along the row, and the intermediate result is expressed as $s=[s_1, s_2, \cdots, s_C]$. Then $s$ is input into a simple feature extracting unit:
\begin{equation}
\widetilde{s}=H_{FEU}\left(s\right)=\sigma(\textbf{L}_{u}\rho(\textbf{L}_{d}s)),
\end{equation}
in which $H_{FEU}(\cdot)$ is composed of two $1\times1$ convolutions and two activation functions. The $\textbf{L}_{d}$ is to map the channel of $s$ to $C/r$, while the other $\textbf{L}_{u}$ is to recover the channel size $C$. $\sigma(\cdot)$ and $\rho(\cdot)$ denote the sigmoid and PReLU function. The former is to normalize the numerical range to $0\sim1$, and the latter aims to boost non-linear adaptability. Finally, the rescaling factor $\widetilde{s}_c$ is to readjust each channel $f_c$ of the input feature:
\begin{equation}
\widetilde{\mathbf{f}}_{c}=\widetilde{s}_c \odot \mathbf{f}_{c},
\end{equation}
where $\widetilde{\mathbf{f}}_{c}$ is the $c$-th channel of refined output $\widetilde{\mathbf{F}}$ and $\odot$ denotes the element-wise product. With the presented TCRM, our HPRN model automatically recalibrates the informative features and enhances the learning power, which can further regularize and optimize the one-to-many mapping space indirectly.

\subsection{Semantic-driven Spatial Relation Module (SSRM)}
\label{SSRM}
%为了便于后续的SSRM描述的作用被理解，这里描述下CNN的特点，并指出缺陷——无法控制相关性不大的特征参与加权，导致表达能力有限。

The principle of correlation learning for CNNs is usually an effort to model similar patterns of spatially-local features via amounts of learnable weights. Since the convolution kernel shape is pre-set, the features involved in correlation expression are restricted to a square neighborhood area. This limitation cannot prevent weakly-correlated features from participating in feature aggregation, nor can it introduce more highly-correlated ones outside the scope, which impedes the learning ability of CNNs. Semantic information is an explicit attribute calibration that converts raw data (such as an RGB image) into a mask with different highlighted regions of interest, where each pixel of the image is clearly assigned as a unique category. The positions of the same identifications naturally tend to have strong correlations and similarities. Generally, local category attributes and scene distributions are consistent between the given RGB images and ground HSIs spatially. Innovatively, we introduce the semantic prior of RGB inputs into SSR, and well-design a SSRM module to fulfill the category-consistent feature aggregation ro refine coarse spectral signatures.

The diagram of SSRM is illustrated in Fig.\ref{figure_ssrm}. Intuitively, the input of SSRM module is constituted of the semantic prior $\mathbf{S}_{RGB}$ of the RGB image and coarse estimated HSI $\mathbf{F}_{Coa}$. $\mathbf{S}_{RGB}$ is obtained through the widely-used simple linear iterative clustering (SLIC) superpixel algorithm \cite{achanta2012slic}. $\mathbf{F}_{Coa}$ comes from the reconstruction part in Section \ref{Network Architecture}. Formally, we seed $\mathbf{F}_{Coa}$ to two $1\times1$ convolutional layers $\varphi(\cdot)$ and $\psi(\cdot)$ to generate two new features $\textbf{D}$ and $\textbf{E}$, respectively, where $\{\mathbf{D}, \mathbf{E}\} \in \mathbb{R}^{B \times H \times W}$. Then $\mathbf{S}_{RGB}$ is embedded into $\mathbf{F}_{Coa}$ as the indexes to unfold and order the features $\{\mathbf{D}, \mathbf{E}\}$ along the $H\times W$ direction:
\begin{equation}
\{\mathbf{D}^{'}, \mathbf{E}^{'}\}=H_{UO}(\{\mathbf{D}, \mathbf{E}\}),
\end{equation}
where $\{\mathbf{D}^{'}, \mathbf{E}^{'}\} \in \mathbb{R}^{B \times (H \times W)}$. Each row of the $\mathbf{D}^{'}$ and $\mathbf{E}^{'}$ is sorted according to the category indexes. To aggregate features of semantically-consistent contents in parallel, we redivide $H\times W$ 1D $B$-length features from $\{\mathbf{D}^{''}, \mathbf{E}^{''}\}$ into $G$ groups spatially, and the results are expressed as $\{\mathbf{D}^{''}, \mathbf{E}^{''}\} \in \mathbb{R}^{B \times G\times N}$. Ideally, the $H\times W$ should be exactly divisible by $G$. When it cannot be satisfied in fact, the mirror filling of boundaries is necessary. For $\{\mathbf{D}^{''}, \mathbf{E}^{''}\}$, the $N$ 1D $B$-length items of most groups belong to the same label. Inevitably, since the feature number of each category is actually unbalanced, the $N$ elements in partial groups contain a few features of the neighbouring category. Next, we reshape and transpose these two features to $\{\mathbf{D}^{''} \in \mathbb{R}^{G \times N\times B}, \mathbf{E}^{''} \in \mathbb{R}^{G \times B\times N}\}$. A semantic-embedded relation matrix $\textbf{Z}$ can be obtained by
\begin{equation}
\mathbf{Z}=\operatorname{softmax}\left({\mathbf{D}^{''}\otimes \mathbf{E}^{''}}\right),
\end{equation}
where $\otimes$ denotes batch matrix multiplication and $\mathbf{Z} \in \mathbb{R}^{G \times N\times N}$. Briefly, this computation broadcasts along the ground dimension. For the $g$-th group $Z_g\in{\mathbb{R}^{N\times N}}$, $N$ weights of its $i$-th row encode the dependence between the $i$-th 1D $B$-length feature of $\mathbf{D}^{''}_g\in{\mathbb{R}^{N\times B}}$ and all the 1D $B$-length features of $\mathbf{E}^{''}_g\in{\mathbb{R}^{B\times N}}$. Similarly, another $1\times1$ convolution $\chi(\cdot)$ is adopted to acquire the new feature $\mathbf{Y}\in \mathbb{R}^{B\times H\times W}$. And the same dimension conversion as $\mathbf{D}$ is implemented to become $\mathbf{Y}^{''} \in \mathbb{R}^{G\times N\times B}$. Finally, the feature aggregation across the clustered similar characteristics is performed by
\begin{equation}
\mathbf{I}_{SSR}=H_{FR}(\mathbf{Z}\otimes \mathbf{Y}^{''}),
\end{equation}
where $H_{FR}(\cdot)$ is the folding and reordering operations, which are the inverse process of $H_{UO}(\cdot)$. At this point, we illustrate detailedly the process of the SSRM with one scale (i.e. the number of categories and Fig. \ref{figure_ssrm} shows $8$ example categories). To reduce the error of once segmentation result, we employ multi-scales SLIC to produce several clustering mappings with different numbers of categories parallelly. Finally, multi-scales SSRM is combined by each single scale result via a $1\times1$ convolution.

\begin{figure}[htbp]
	\centering
	\subfloat[]{\includegraphics[width=2.1cm,height=3.8cm]{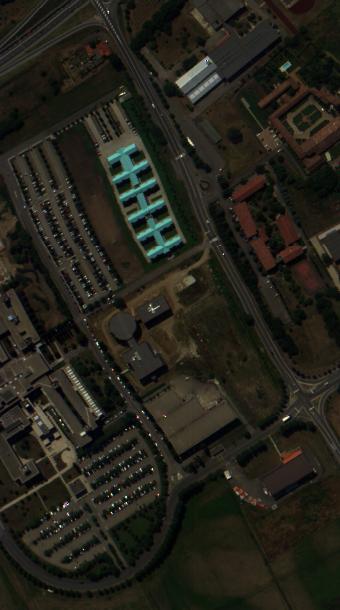}}\
	\subfloat[]{\includegraphics[width=2.1cm,height=3.8cm]{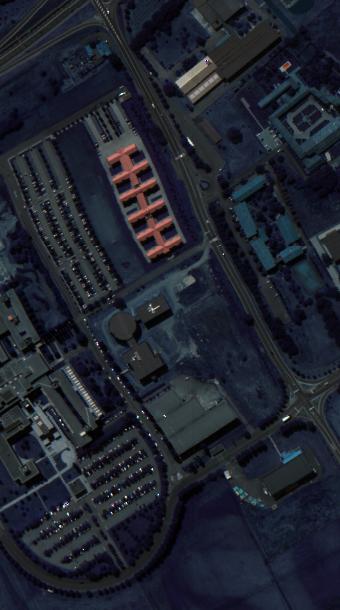}}\
	\subfloat[]{\includegraphics[width=2.1cm,height=3.8cm]{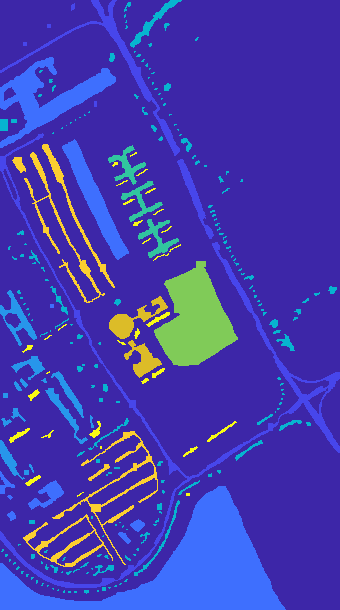}}\
	\subfloat[]{\includegraphics[width=2.1cm,height=3.8cm]{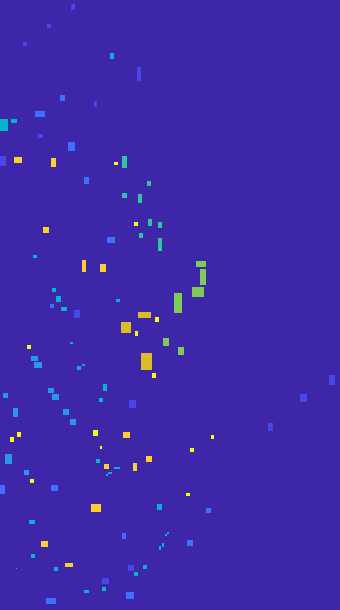}}\\
	\caption{UP dataset. (a) Simulated RGB input. (b) False-color image (25th, 45th, and 70th bands). (c) Label. (d) Standard training samples.}
	\label{figure_pavia_intro}
\end{figure}

\subsection{Second-Order Prior Constraint (SOPC)}
Second-order statistics plays a critical role to explore the correlations of different samples \cite{li2020adaptive, dai2019second}. From the imaging principle of HSIs, there are mathematical correlations among different bands and spectral consistency along the band direction. Take the reconstructed spectral $\mathbf{I}_{SSR}\in \mathbb{R}^{B \times H\times W}$ as an example, we reshape it as $\mathbf{I}_{SSR}\in \mathbb{R}^{B \times (H\times W)}$, which means that there are $B$ samples of $n=H\times W$ length. Then the sample covariance matrix can be calculated by
\begin{equation}
\mathbf{X}_{SSR}=\mathbf{I}_{SSR}\otimes \overline{\boldsymbol{\Sigma}}\otimes \mathbf{I}_{SSR}^{\mathrm{T}},
\end{equation}
where $\overline{\boldsymbol{\Sigma}}=\frac{1}{n}\left(\boldsymbol{\Sigma}-\frac{1}{n} \mathbf{1}\right)$. $\boldsymbol{\Sigma}$ and $\mathbf{1}$ are $n\times n$ identity matrix and matrix of all ones. The element in the $i$-th row and $j$-th column of matrix $\mathbf{X}_{SSR}\in\mathbb{R}^{B \times B}$ represents the correlation between the $i$-th band and the $j$-th one for $\mathbf{I}_{SSR}$. In the same way, we can get the normalized covariance matrix $\mathbf{X}_{HSI}$ of ground HSI $\mathbf{I}_{HSI}$. In order to maintain the mathematical correlation and spectral consistency between hyperspectral bands, the second-order prior constraints (SOPC) are incorporated into the loss function:
\begin{equation}
L(\Theta)=\left\|\mathbf{I}_{HSI}-\mathbf{I}_{SSR}\right\|_{1}+\tau\left\|\mathbf{X}_{HSI}-\mathbf{X}_{SSR}\right\|_{1},
\end{equation}
where $\Theta$ denotes the parameter set of our HPRN, and $\tau$ is the trade-off weight. In our experimental settings, the trade-off value was experimentally tried with $\{0.2, 0.5, 1, 2, 5, 10\}$, and finally determined to be $2$ using the NTIRE2018 validation set. From the mathematical perspective, the SOPC term can help L1 loss to compress the space of the possible one-to-many mapping function to achieve a high-precision spectral recovery.

\begin{table}[]
	\centering
	\renewcommand\arraystretch{1.2}
	\caption{The detailed dataset division of NTIRE2018 ``Clean'' and ``Real World'' tracks.}
	\setlength{\tabcolsep}{4mm}{
		\begin{tabular}{@{}ccc@{}}
			\toprule[1.4pt]
			Our testing set & Our verification set &Our training set\\ \midrule[1.4pt]
			\begin{tabular}[c]{@{}c@{}}
				257,259,261,263,265.
			\end{tabular} & \begin{tabular}[c]{@{}c@{}}
				001,036,204,209,225.
			\end{tabular} & \begin{tabular}[c]{@{}c@{}}The remain of the \\ official training set.\end{tabular}  \\ \bottomrule[1.4pt]
	\end{tabular}}
	\label{table_2018_division}
\end{table}

\begin{table}[]
	\centering
	\renewcommand\arraystretch{1.2}
	\caption{The detailed dataset division of NTIRE2020 ``Clean'' and ``Real World'' tracks.}
	\setlength{\tabcolsep}{4mm}{
		\begin{tabular}{@{}ccc@{}}
			\toprule[1.4pt]
			Our testing set & Our verification set &Our training set\\ \midrule[1.4pt]
			\begin{tabular}[c]{@{}c@{}}
			451,453,455,456,457,\\459,462,463,464,465.
		    \end{tabular} & \begin{tabular}[c]{@{}c@{}}
		    079,089,255,304,363,\\372,387,422,434,446.
	    \end{tabular} & \begin{tabular}[c]{@{}c@{}}The remain of the \\ official training set.\end{tabular}  \\ \bottomrule[1.4pt]
	\end{tabular}}
	\label{table_2020_division}
\end{table}

\section{Experiments}
\subsection{Experimental Setting}
\subsubsection{SSR Benchmark Datasets} The presented HPRN is evaluated on four public datasets, ie. both ``Clean'' and ``Real World'' tracks of two SSR challenges NTIRE2018 \cite{arad2018ntire} and NTIRE2020 \cite{arad2020ntire}. For ``Clean'' tracks, the HSIs are estimated from noise-free RGB images, which are calculated numerically using the ground-truth HSI and given spectral sensitivity functions. ``Real World'' tracks simulate capturing by an uncalibrated and unknown camera, where the HSIs are recovered from noisy JPEG-compressed RGB images. Certainly, the camera response functions for the same tracks of the NTIRE2020 are changed over the NTIRE2018. A Specim PS Kappa DX4 hyperspectral camera is used for collecting NTIRE2018 datasets with a rotary stage for spatial scanning, while NTIRE2020 datasets are acquired by a stand-alone, battery-powered, push-broom spectral imaging system equipped with a Specim IQ mobile hyperspectral camera. The NTIRE2018 dataset consists of 256 training RGB-HSI pairs, 5 validating RGB-HSI pairs and 10 testing RGB inputs. The HSIs contain 31 spectral bands (400-700nm at roughly 10nm increments) with ${1392 \times 1300}$ size. Also, there are 450 training RGB-HSI pairs, 10 validating RGB-HSI pairs and 20 testing RGB images. All HSIs possess the size of ${512 \times 482}$ with 31 bands from 400nm to 700nm at 10nm steps. Due to the unavailable official testing spectra, we decide to take the official verification set as our test set, and randomly choose several images from the official training as our verification set in this paper. The rest of the official training is adopted as our training set. The specific description refers to Table \ref{table_2018_division} and Table \ref{table_2020_division}.
  
\subsubsection{Remote Sensing Dataset} The University of Pavia (UP) dataset is chosen to verify the effectiveness of the reconstructed spectrum based on classification results \cite{xu2017multisource}. The Reflective Optics System Imaging Spectrometer (ROSIS) sensor is employed to collect this data over the area of Pavia, Northern Italy. The scene has 103 spectral bands with a range of 430$\sim$860nm, and the spatial resolution is 1.3m with the size $610\times340$. There are about 43923 samples labeled with nine classes. According to the reference \cite{xu2017multisource}, the $53$rd, $31$st, and $7$th bands of the original HSI are selected to form as the simulated RGB data. The simulated RGB input, the false-color image of the HSI, the groundtruth label and the standard training samples are drawn in Fig.\ref{figure_pavia_intro}(a), (b), (c) and (d), respectively. The corresponding testing set is the remaining part of the groundtruth label except the training samples.

\subsubsection{Quantitative Metrics} To evaluate the performance of HPRN quantitatively, five metrics including the mean relative absolute error (MRAE), root mean square error (RMSE), spectral angle mapper (SAM), peak signal-to-noise ratio (PSNR), and average structural similarity (ASSIM) are utilized:
\begin{equation}
	\operatorname{MRAE}=\frac{1}{\mathbb{N}} \sum_{p=1}^{\mathbb{N}}\left(\left|\textbf{I}_{HSI}^{(p)}-\textbf{I}_{SSR}^{(p)}\right| / \textbf{I}_{HSI}^{(p)}\right),
\end{equation}
\begin{equation}
	\operatorname{RMSE}=\sqrt{\frac{1}{\mathbb{N}} \sum_{p=1}^{\mathbb{N}}\left(\textbf{I}_{HSI}^{(p)}-\textbf{I}_{SSR}^{(p)}\right)^{2}},
\end{equation}
\begin{equation}
	\operatorname{SAM}=\frac{1}{\mathbb{M}} \sum_{v=1}^{\mathbb{M}}\left(\frac{arccos(<\textbf{I}_{HSI}^{(v)}, \textbf{I}_{SSR}^{(v)}>}{ (||\textbf{I}_{HSI}^{(v)}||_2 || \textbf{I}_{SSR}^{(v)}||_2))} \right),
\end{equation}
\begin{equation}
	\operatorname{PSNR}=-\frac{10}{\mathbb{N}} \sum_{p=1}^{\mathbb{N}} \log \left(\textbf{I}_{HSI}^{(p)}-\textbf{I}_{SSR}^{(p)}\right)^{2},
\end{equation}
\begin{equation}
	\operatorname{ASSIM}=\frac{1}{B} \sum_{b=1}^{B} \operatorname{SSIM}\left(\textbf{I}_{HSI}^{(b)}, \textbf{I}_{SSR}^{(b)}\right),
\end{equation}
where ${\textbf{I}_{SSR}^{(p)}}$ and ${\textbf{I}_{HSI}^{(p)}}$ represent the ${p}$-th pixel of the super-resolved and groundtruth HSI. $<\textbf{I}_{HSI}^{(v)}, \textbf{I}_{SSR}^{(v)}>$ are the dot product of the two ${v}$-th spectral vectors between the groundtruth and the estimated spectrum. $||\cdot||$ denotes $l2$ norm function. ${\mathbb{N}}$, ${\mathbb{M}}$ and ${\mathbb{B}}$ are the number of pixels, spectral vectors and bands of the HSI cube, respectively. $\operatorname{SSIM}({\cdot})$ calculates the SSIM value of a
typical band. Mathematically, the smaller the MRAE, RMSE and SAM, the better predicted HSIs. Moreover, the larger the PSNR and SSIM, the less errors between the reconstructed result and groundtruth.

\begin{table}[]
	\centering
	\renewcommand\arraystretch{1.2}
	\caption{Explore the location of the TCRM with the squeezed vector size $4\times4$.}
	\setlength{\tabcolsep}{12mm}{
	\begin{tabular}{@{}lcc@{}}
		\toprule[1.4pt]
		Description        & MRAE(↓) & SAM(↓) \\ \midrule[1.4pt]
		\textit{baseline}  & 0.01538 & 1.089  \\
		\textit{1-pos}     & 0.01493 & 1.036  \\
		\textit{2-pos}     & 0.01501 & 1.049  \\
		\textit{3-pos}     & 0.01487 & 1.020  \\
		\textit{multi-pos} & 0.01481 & 1.011  \\ \bottomrule[1.4pt]
	\end{tabular}}
\label{table_tcrm_pos}
\end{table}

\begin{table}[]
	\centering
	\renewcommand\arraystretch{1.2}
	\caption{Explore the squeezed vector size of the TCRM.}
	\setlength{\tabcolsep}{2.96mm}{
	\begin{tabular}{@{}lcccc@{}}
		\toprule[1.4pt]
		Description & MRAE(↓)   & SAM(↓)  & Params(M) & MACs(K) \\ \midrule[1.4pt]
		\textit{baseline}        & 0.01538 & 1.089 & ———      & ———    \\
		$1\times1$       & 0.01516 & 1.051 & 8.25     & 48.04   \\
		$2\times2$          & 0.01504 & 1.030 & 8.38     & 48.64   \\
		$4\times4$          & 0.01487 & 1.020 & 10.3     & 58.24   \\
		$8\times8$         & 0.01504 & 1.042 & 41.1     & 218.1   \\ \bottomrule[1.4pt]
	\end{tabular}}
\label{table_tcrm_size}
\end{table}

\begin{table}[]
	\centering
	\renewcommand\arraystretch{1.2}
	\caption{Effects of sharing weights of embedding functions for the SSRM.}
	\setlength{\tabcolsep}{11.6mm}{
	\begin{tabular}{@{}lcc@{}}
		\toprule[1.4pt]
		Description        & MRAE(↓) & SAM(↓) \\ \midrule[1.4pt]
		\textit{baseline}  & 0.01538 & 1.089  \\
		$\varphi(\cdot)=\psi(\cdot)$     & 0.01494 & 1.055  \\
		$\varphi(\cdot)\neq\psi(\cdot)$  & 0.01514 & 1.067  \\ \bottomrule[1.4pt]
\end{tabular}}
	\label{table_ssrm_share}
\end{table}

\begin{figure}
	\centering
	\includegraphics[width=0.47\textwidth]{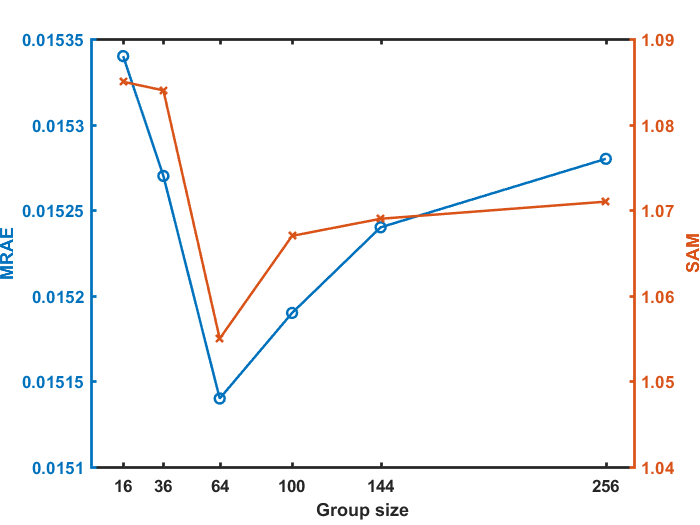}
	\caption{Effects of the group size of the SSRM.}
	\label{figure_ssrm_group}
\end{figure}

\begin{table}[]
	\centering
	\renewcommand\arraystretch{1.2}
	\caption{Effects of the scales settings of the SSRM.}
	\setlength{\tabcolsep}{12mm}{
		\begin{tabular}{@{}lcc@{}}
			\toprule[1.4pt]
			Description        & MRAE(↓) & SAM(↓) \\ \midrule[1.4pt]
			\textit{baseline}  & 0.01538 & 1.089  \\
			\textit{4,6,8,10}     & 0.01514 & 1.055  \\
			\textit{8,12,16,20}     & 0.01494 & 1.055  \\
			\textit{12,18,24,30}     & 0.01520 & 1.073  \\
			\textit{16,24,32,40} & 0.01519 & 1.082  \\ \bottomrule[1.4pt]
	\end{tabular}}
	\label{table_ssrm_scale}
\end{table}

\begin{table}[]
	\centering
	\renewcommand\arraystretch{1.2}
	\caption{Effectiveness of different modules.}
	\setlength{\tabcolsep}{3mm}{
		\begin{tabular}{@{}lccccc@{}}
			\toprule[1.4pt]
			Description        & \textit{SSRM} & \textit{TCRM}& \textit{SOPC} &MRAE(↓)& SAM(↓) \\ \midrule[1.4pt]
			\textit{baseline}  & $\times$ &$\times$&$\times$& 0.01538 & 1.089  \\
			\textit{with SSRM}     & $\checkmark$ &$\times$&$\times$ & 0.01494 & 1.055  \\
			\textit{with TCRM}     &$\times$&$\checkmark$&$\times$& 0.01487 & 1.020  \\
			\textit{with SOPC}     &$\times$&$\times$&$\checkmark$& 0.01520 & 1.075  \\
			\textit{w/o TCRM}     &$\checkmark$&$\times$&$\checkmark$& 0.01484 & 1.046\\
			\textit{w/o SSRM}     &$\times$&$\checkmark$&$\checkmark$& 0.01478 & 1.017  \\
			\textit{w/o SOPC}     &$\checkmark$&$\checkmark$&$\times$& 0.01455 & 1.018\\
			\textit{HPRN} &$\checkmark$&$\checkmark$&$\checkmark$&0.01441 & 1.009\\ \bottomrule[1.4pt]
	\end{tabular}}
	\label{table_hprn}
\end{table}

\subsubsection{Implementation Detail}
For the NTIRE2020 and NTIRE2018 datasets, $64\times64$ patches are cropped from the original RGB-HSI counterparts in the training procedure. The ADAM \cite{kingma2014adam} is adopted as the optimizer of our HPRN. The exponential decay rates are set as ${\beta_1 = 0.9}$ and ${\beta_2 = 0.99}$ for the first and second moment estimates, respectively. The learning rate is set to $0.00012$, and the polynomial function is applied as the decay means with power $=1.5$. As for the backbone framework, there are $M=10$ MRBs, where all intermediate features have $200$ channels. The parameter $r$ in TCRM is $16$ and the group size $g=64$ in SSRM. Besides, the best results are reported within $100$ epochs, and the proposed HPRN network has been implemented on the Pytorch framework via an NVIDIA 2080Ti GPU. The package of automatic mixed precision is also employed to speed up the training of the network. For the remote sensing UP dataset, we arrange the $11\times11$ neighbor patch to incorporate the spectral-spatial statistics according to the positions of training and testing samples, following the general setting of the CNN-based HSI classification method \cite{xu2017multisource}.

\begin{figure*}[!tbp]
	%\vskip -0.05in
	\centering
	\scalebox{1}
	{
		\begin{tabular}{@{}c@{}c@{}c@{}c@{}c@{}c@{}c@{}c@{}c}
			&$\text{Galliani\cite{galliani2017learned}}$ &$\text{MSCNN\cite{yan2018accurate}}$&$\text{UNet\cite{stiebel2018reconstructing}}$&$\text{HSCNN+\cite{shi2018hscnn+}}$&$\text{FMNet\cite{zhang2020pixel}}$& $\text{HRNet\cite{zhao2020hierarchical}}$ &$\text{Ours}$ \\
			\rotatebox{90}{\ \ \ \ \ \
				$\text{Spectra}$}\ \ 
			&{\includegraphics[width=2.42cm,height=2.42cm]{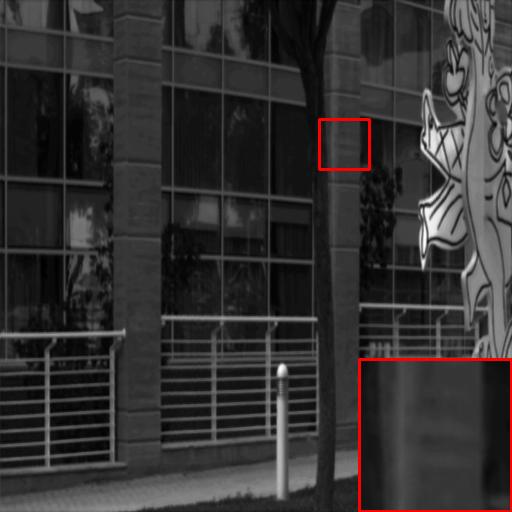}} \
			&{\includegraphics[width=2.42cm,height=2.42cm]{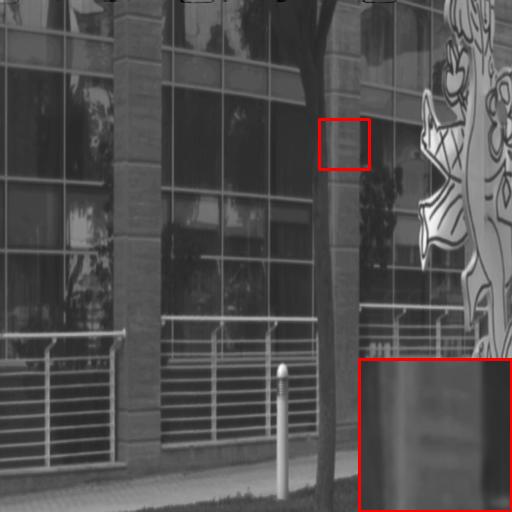}} \
			&{\includegraphics[width=2.42cm,height=2.42cm]{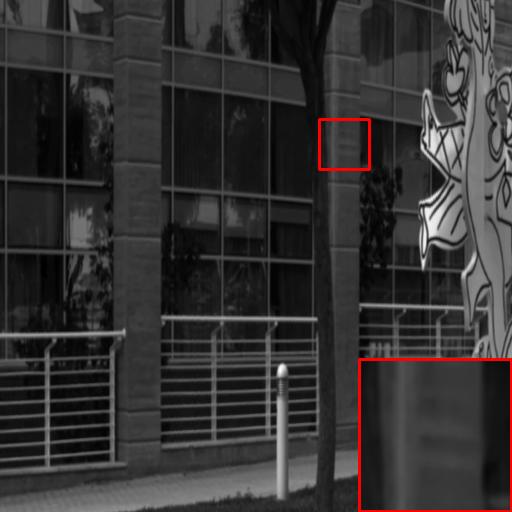}} \
			&{\includegraphics[width=2.42cm,height=2.42cm]{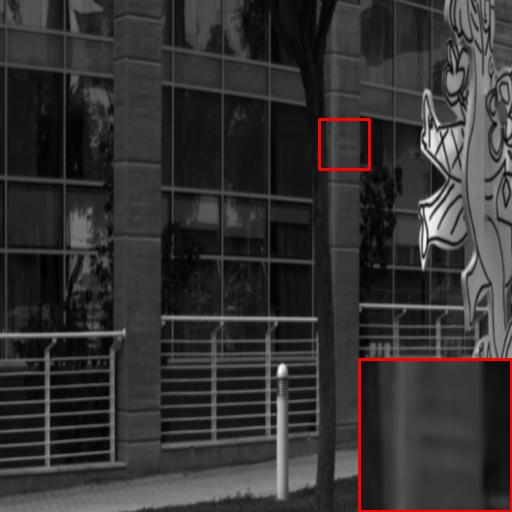}} \
			&{\includegraphics[width=2.42cm,height=2.42cm]{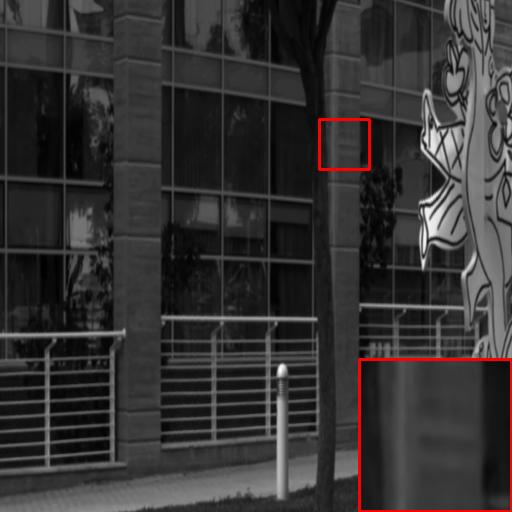}} \
			&{\includegraphics[width=2.42cm,height=2.42cm]{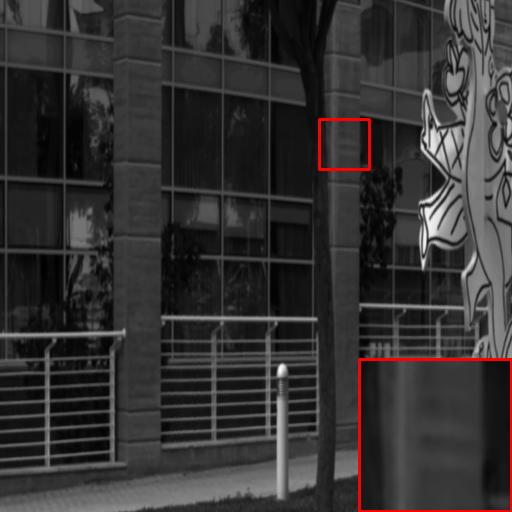}} \
			&{\includegraphics[width=2.42cm,height=2.42cm]{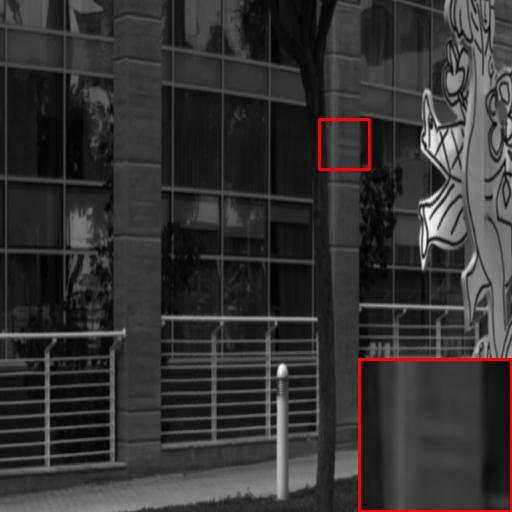}} \\
			\rotatebox{90}{\ \ \ \ \ \ \ 
				$\text{Error}$}\ \ 
			&{\includegraphics[width=2.42cm,height=2.42cm]{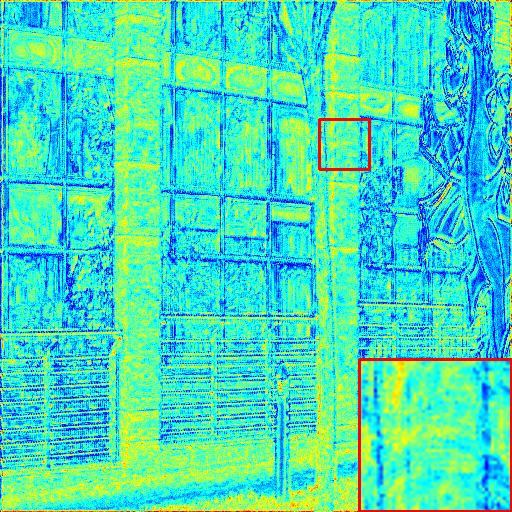}} \
			&{\includegraphics[width=2.42cm,height=2.42cm]{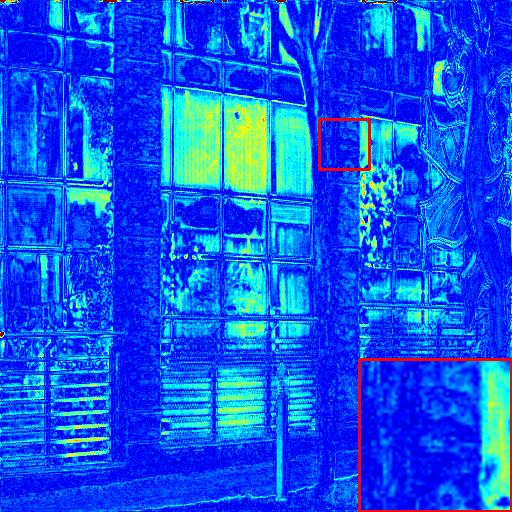}} \
			&{\includegraphics[width=2.42cm,height=2.42cm]{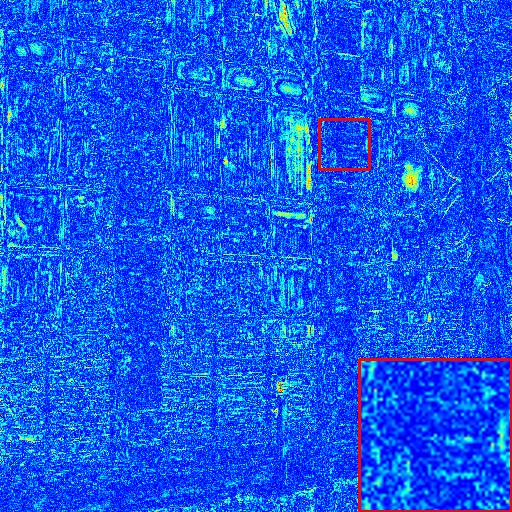}} \
			&{\includegraphics[width=2.42cm,height=2.42cm]{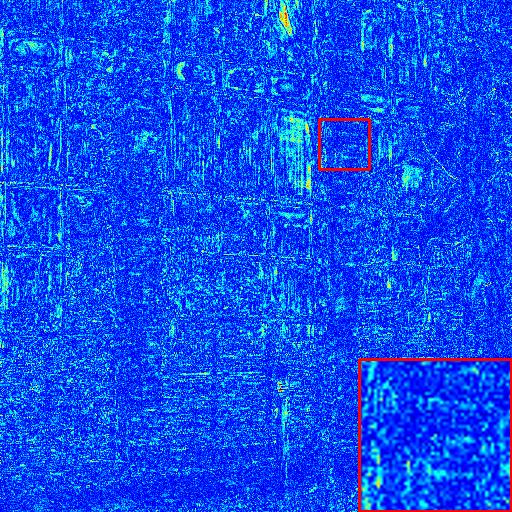}} \
			&{\includegraphics[width=2.42cm,height=2.42cm]{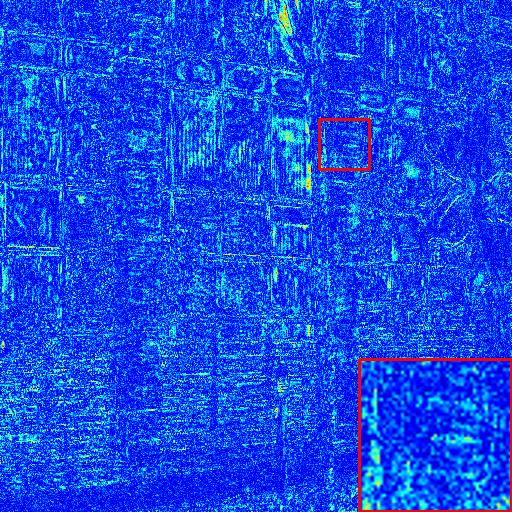}} \
			&{\includegraphics[width=2.42cm,height=2.42cm]{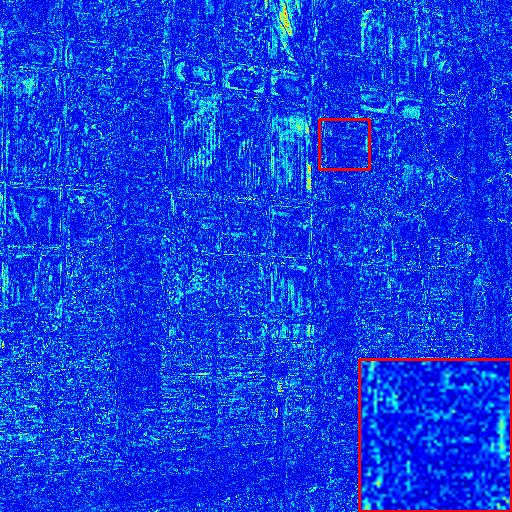}} \
			&{\includegraphics[width=2.42cm,height=2.42cm]{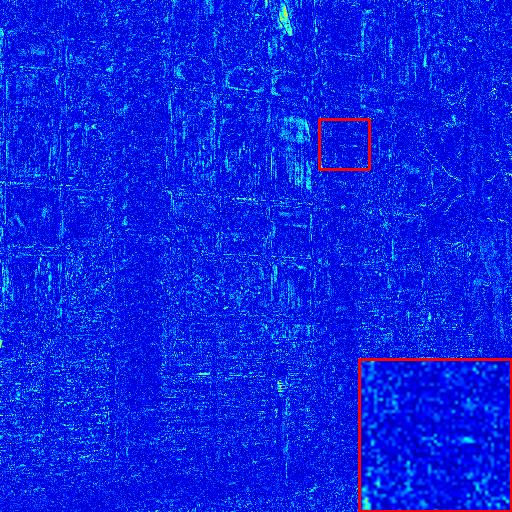}} \\
			\rotatebox{90}{\ \ \ \ \ \
				$\text{Spectra}$}\ \ 
			&{\includegraphics[width=2.42cm,height=2.42cm]{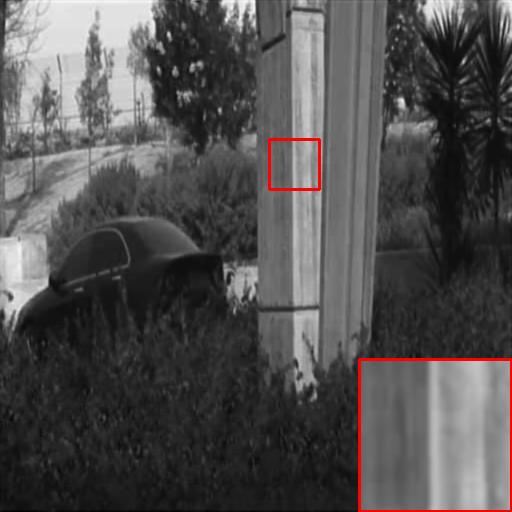}} \
			&{\includegraphics[width=2.42cm,height=2.42cm]{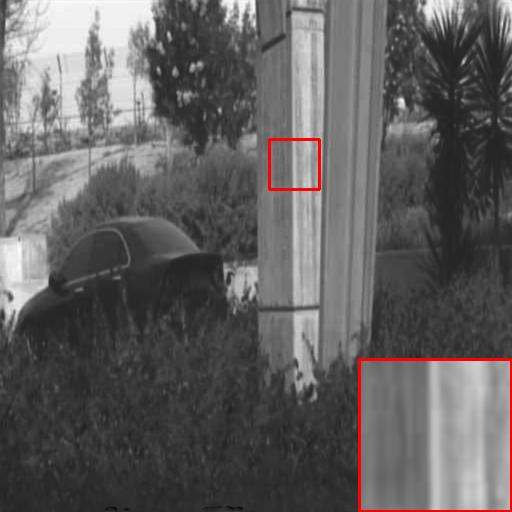}} \
			&{\includegraphics[width=2.42cm,height=2.42cm]{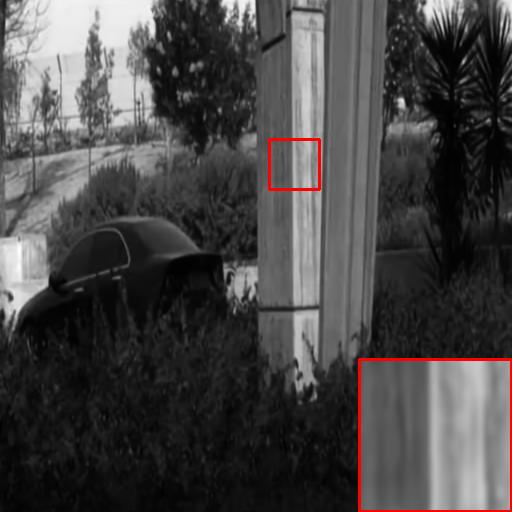}} \
			&{\includegraphics[width=2.42cm,height=2.42cm]{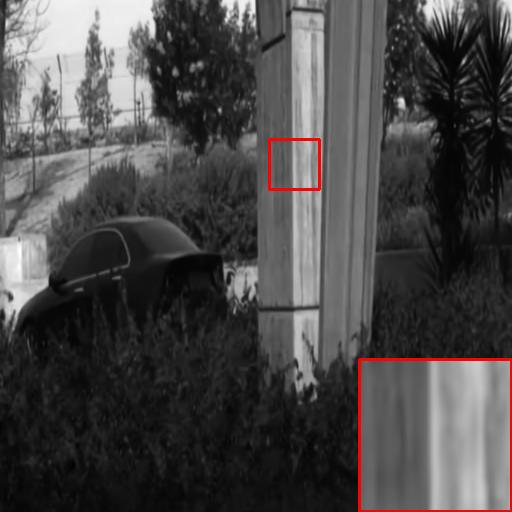}} \
			&{\includegraphics[width=2.42cm,height=2.42cm]{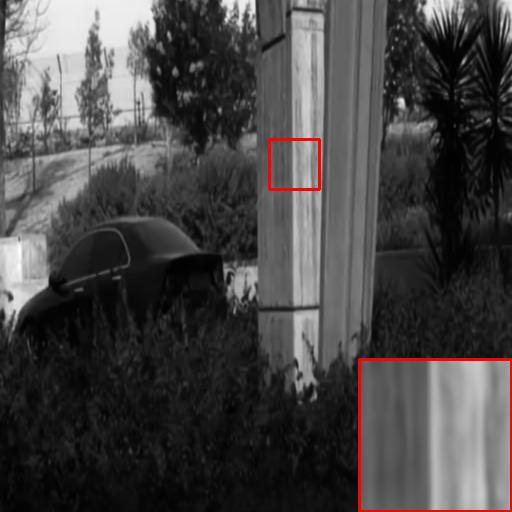}} \
			&{\includegraphics[width=2.42cm,height=2.42cm]{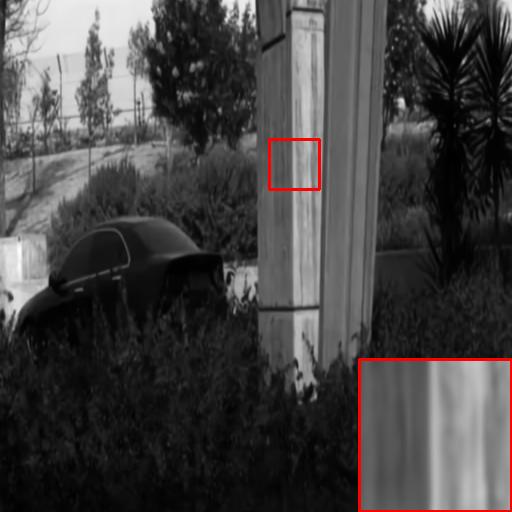}} \
			&{\includegraphics[width=2.42cm,height=2.42cm]{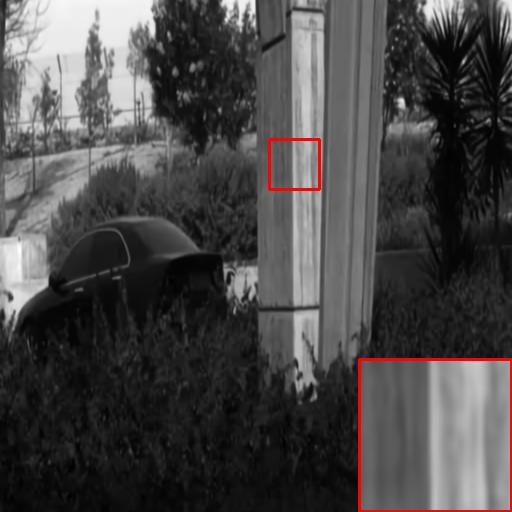}} \\
			\rotatebox{90}{\ \ \ \ \ \ \ 
				$\text{Error}$}\ \ 
			&{\includegraphics[width=2.42cm,height=2.42cm]{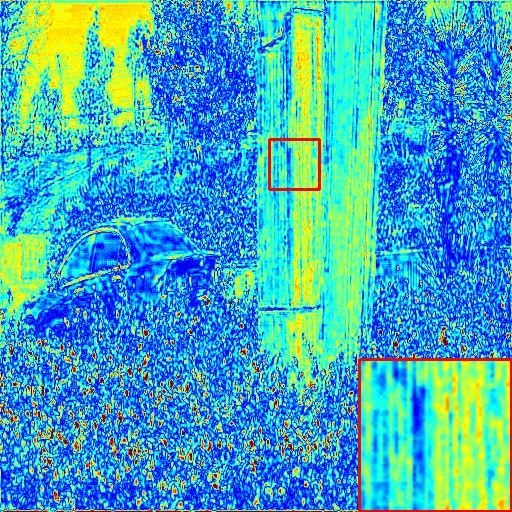}} \
			&{\includegraphics[width=2.42cm,height=2.42cm]{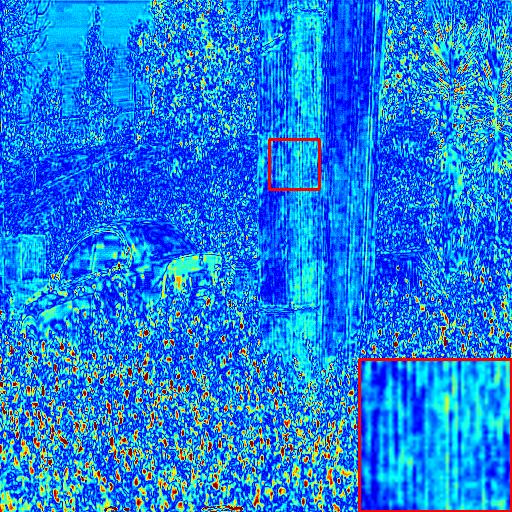}} \
			&{\includegraphics[width=2.42cm,height=2.42cm]{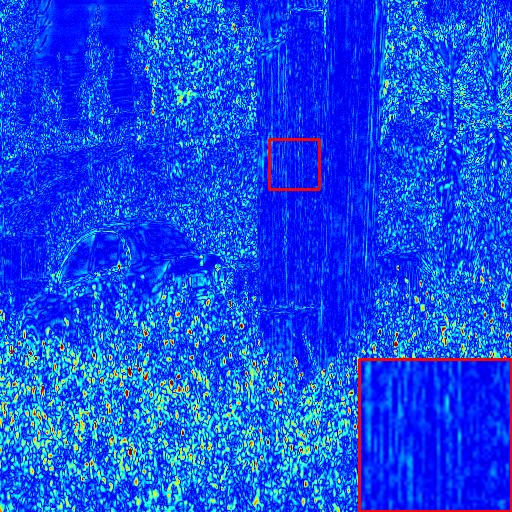}} \
			&{\includegraphics[width=2.42cm,height=2.42cm]{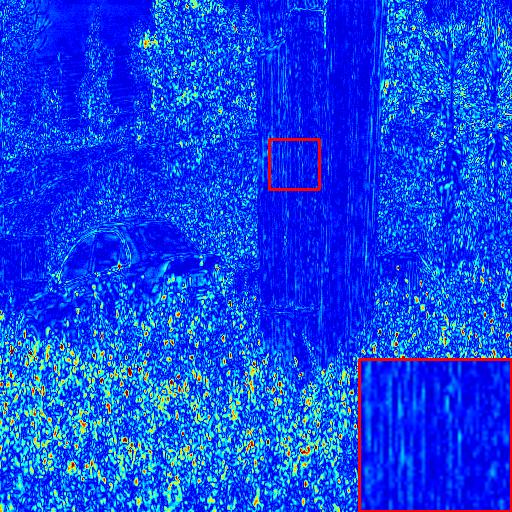}} \
			&{\includegraphics[width=2.42cm,height=2.42cm]{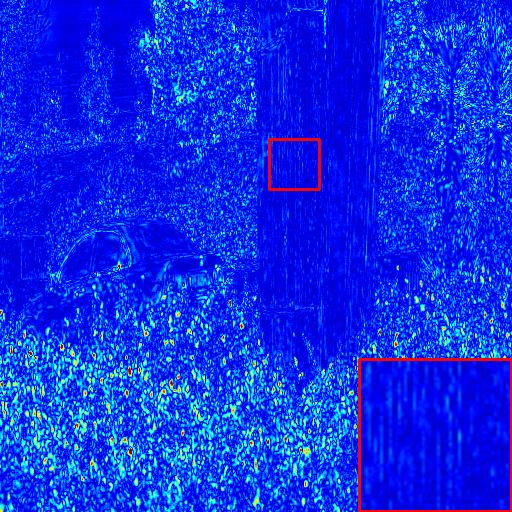}} \
			&{\includegraphics[width=2.42cm,height=2.42cm]{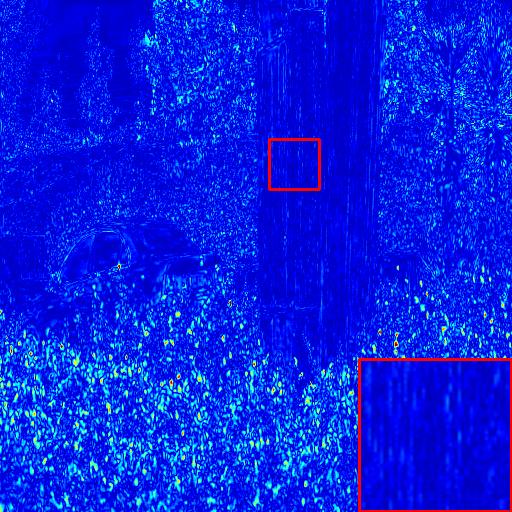}} \
			&{\includegraphics[width=2.42cm,height=2.42cm]{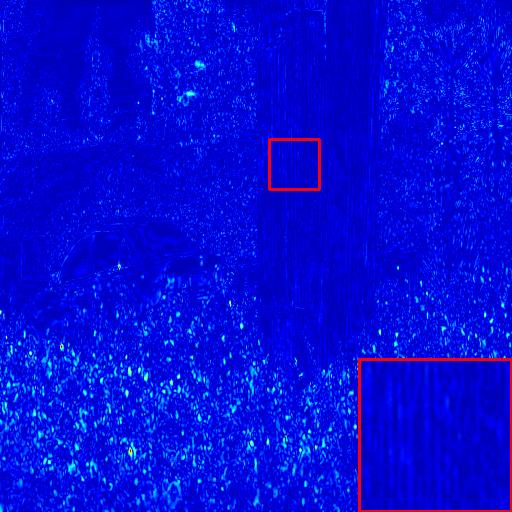}} \\
			\rotatebox{90}{\ \ \ \ \ \
				$\text{Spectra}$}\ \ 
			&{\includegraphics[width=2.42cm,height=2.42cm]{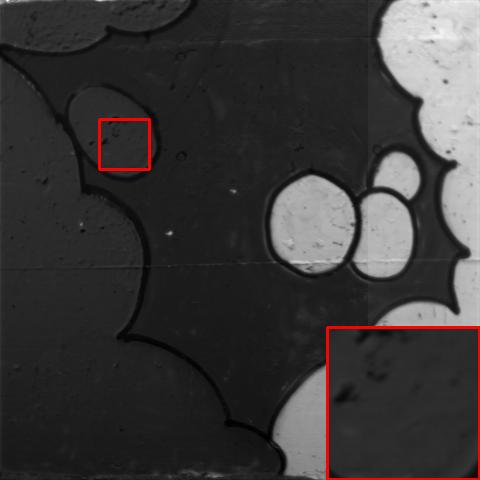}} \
			&{\includegraphics[width=2.42cm,height=2.42cm]{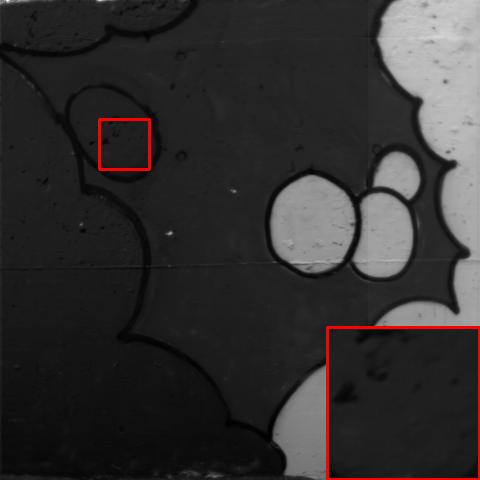}} \
			&{\includegraphics[width=2.42cm,height=2.42cm]{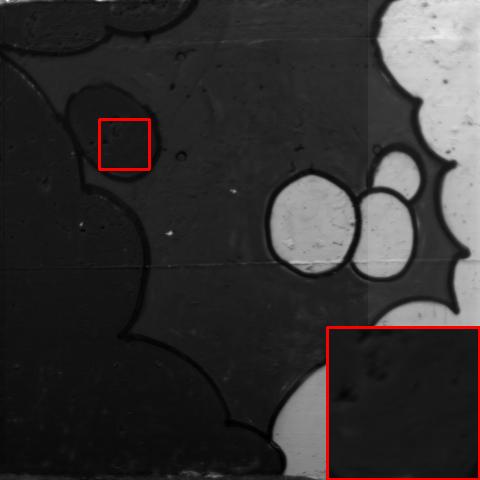}} \
			&{\includegraphics[width=2.42cm,height=2.42cm]{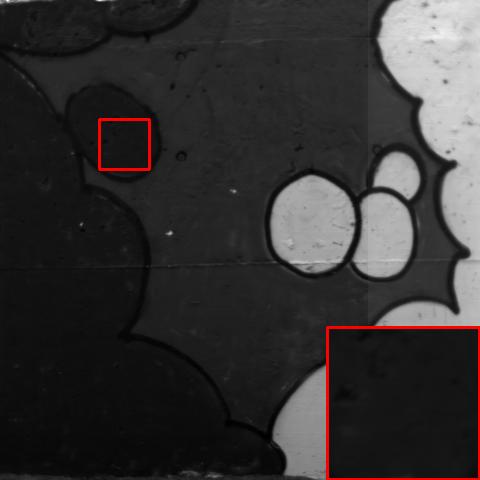}} \
			&{\includegraphics[width=2.42cm,height=2.42cm]{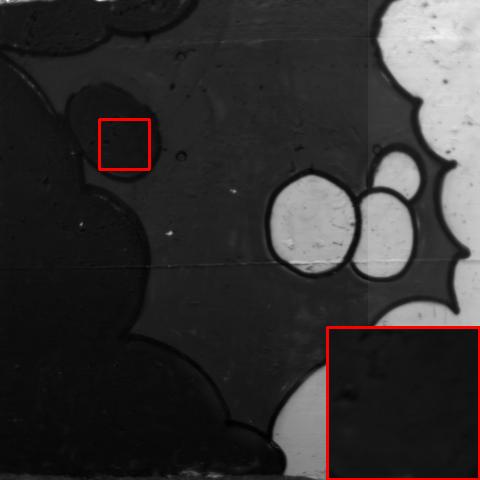}} \
			&{\includegraphics[width=2.42cm,height=2.42cm]{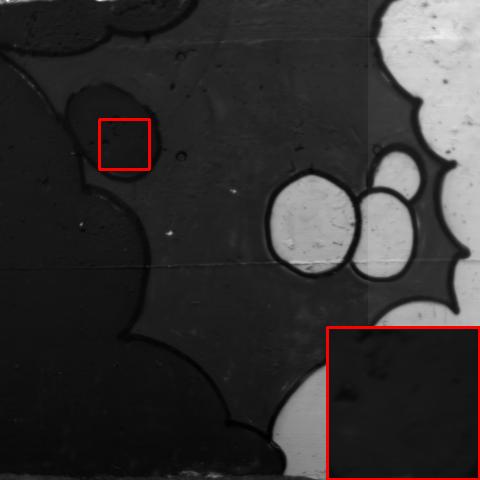}} \
			&{\includegraphics[width=2.42cm,height=2.42cm]{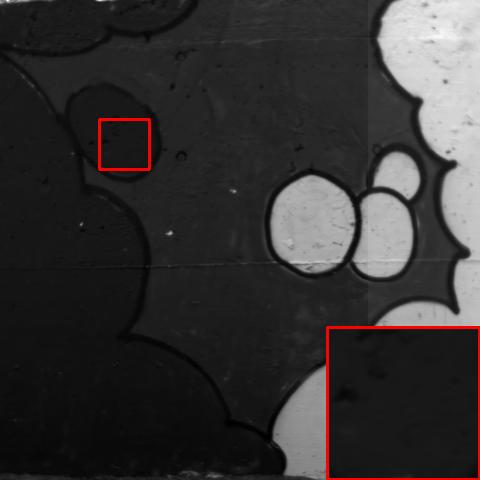}} \\
			\rotatebox{90}{\ \ \ \ \ \ \ 
				$\text{Error}$}\ \ 
			&{\includegraphics[width=2.42cm,height=2.42cm]{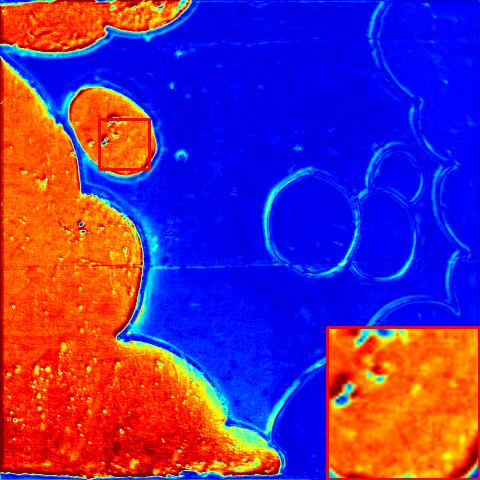}} \
			&{\includegraphics[width=2.42cm,height=2.42cm]{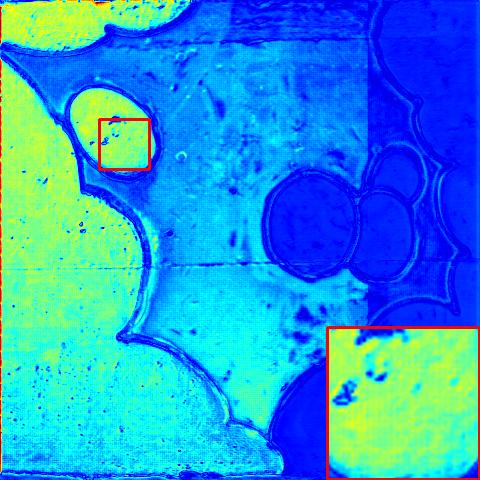}} \
			&{\includegraphics[width=2.42cm,height=2.42cm]{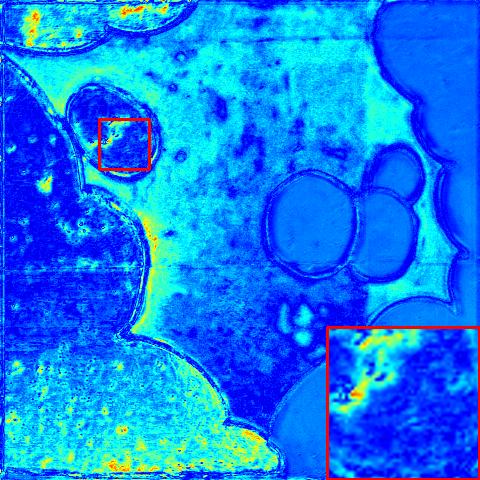}} \
			&{\includegraphics[width=2.42cm,height=2.42cm]{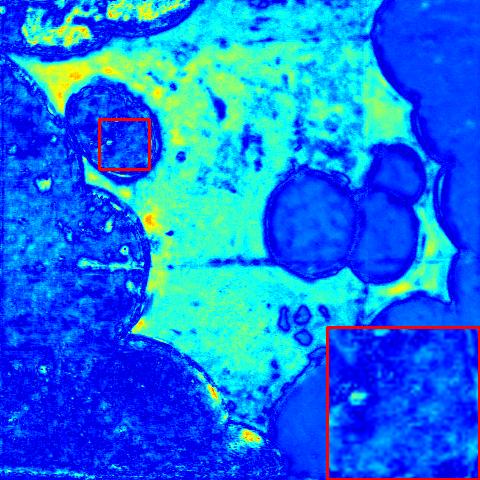}} \
			&{\includegraphics[width=2.42cm,height=2.42cm]{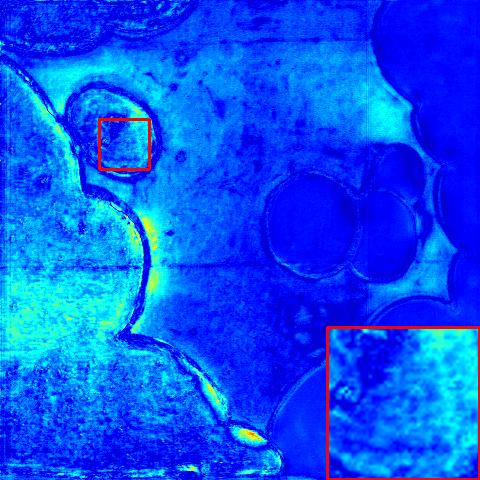}} \
			&{\includegraphics[width=2.42cm,height=2.42cm]{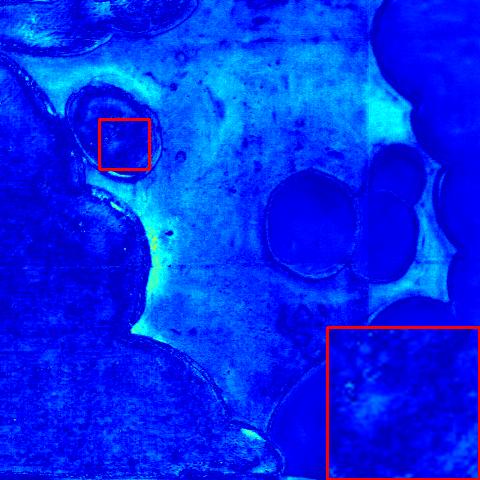}} \
			&{\includegraphics[width=2.42cm,height=2.42cm]{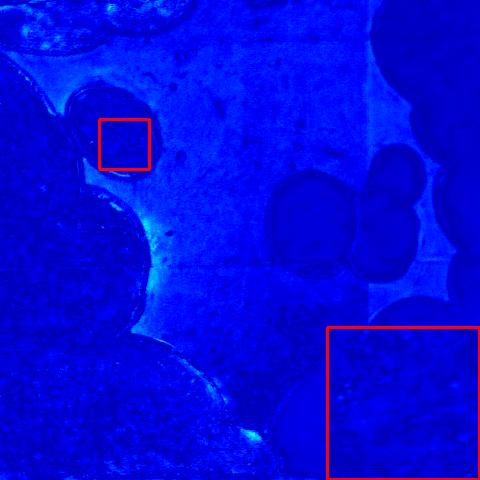}} \\
			\rotatebox{90}{\ \ \ \ \ \
				$\text{Spectra}$}\ \ 
			&{\includegraphics[width=2.42cm,height=2.42cm]{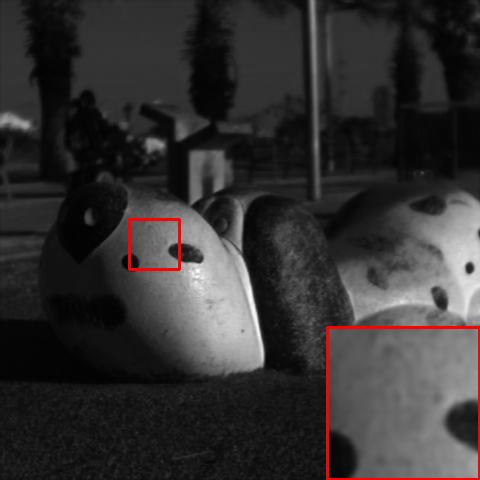}} \
			&{\includegraphics[width=2.42cm,height=2.42cm]{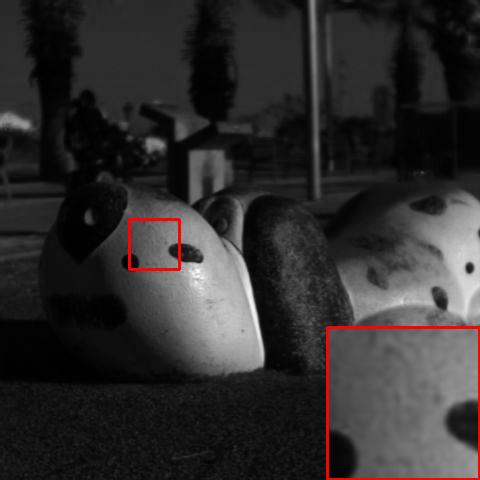}} \
			&{\includegraphics[width=2.42cm,height=2.42cm]{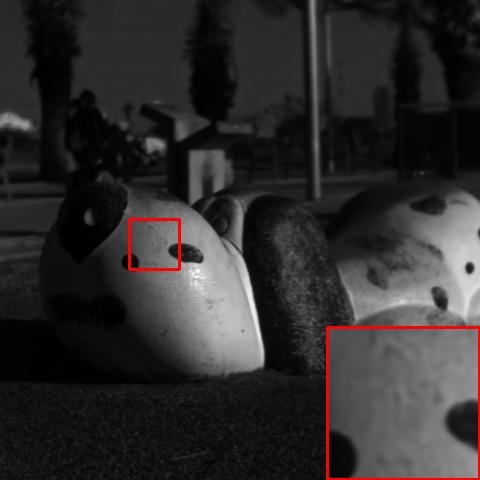}} \
			&{\includegraphics[width=2.42cm,height=2.42cm]{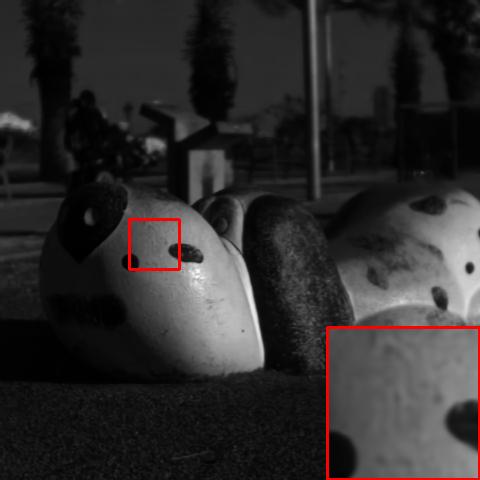}} \
			&{\includegraphics[width=2.42cm,height=2.42cm]{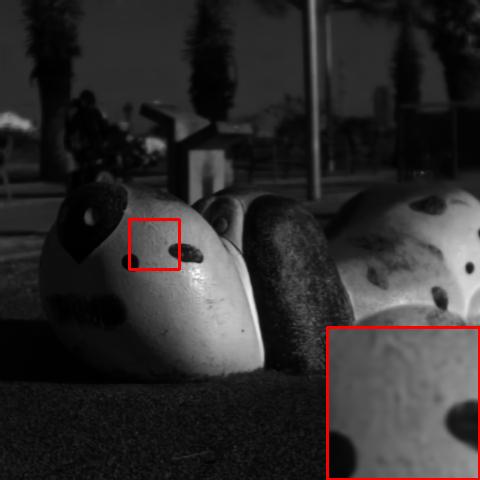}} \
			&{\includegraphics[width=2.42cm,height=2.42cm]{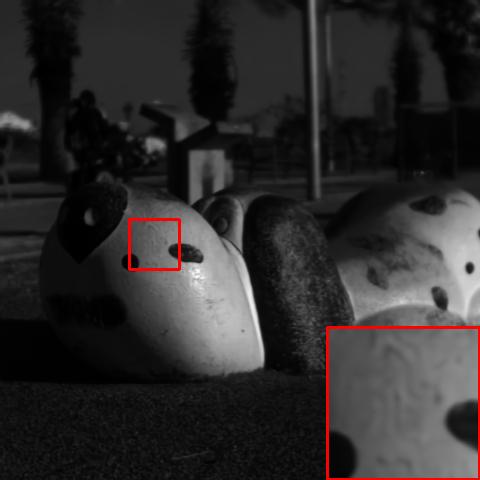}} \
			&{\includegraphics[width=2.42cm,height=2.42cm]{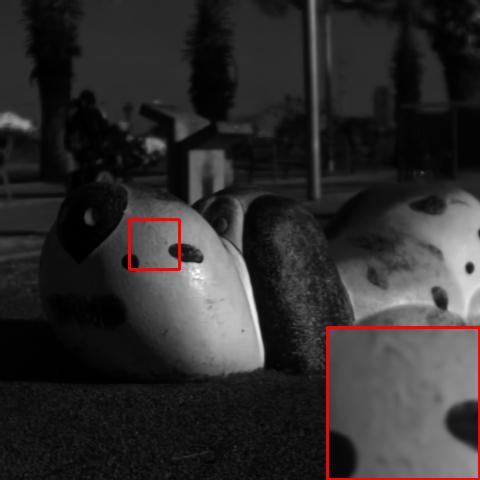}} \\
			\rotatebox{90}{\ \ \ \ \ \ \ 
				$\text{Error}$}\ \ 
			&{\includegraphics[width=2.42cm,height=2.42cm]{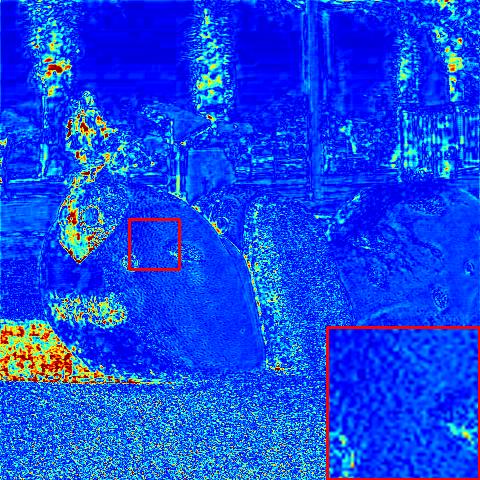}} \
			&{\includegraphics[width=2.42cm,height=2.42cm]{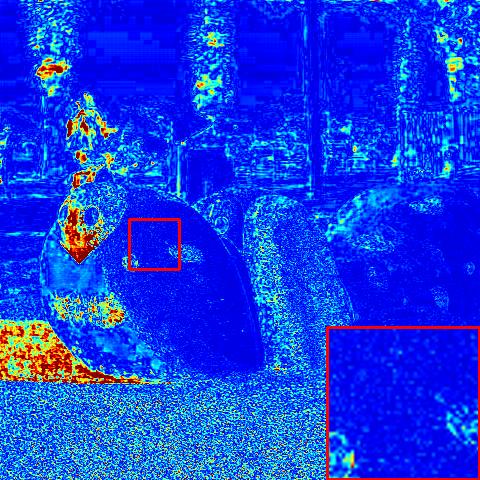}} \
			&{\includegraphics[width=2.42cm,height=2.42cm]{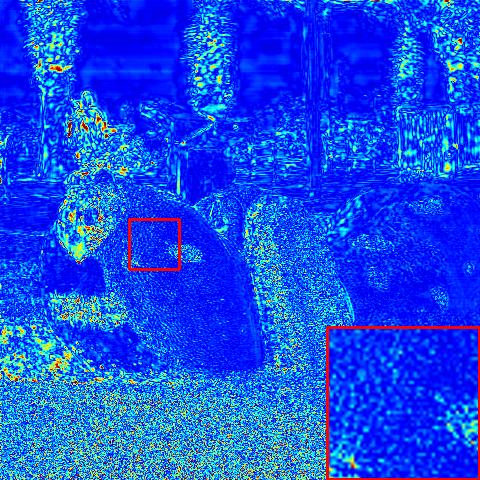}} \
			&{\includegraphics[width=2.42cm,height=2.42cm]{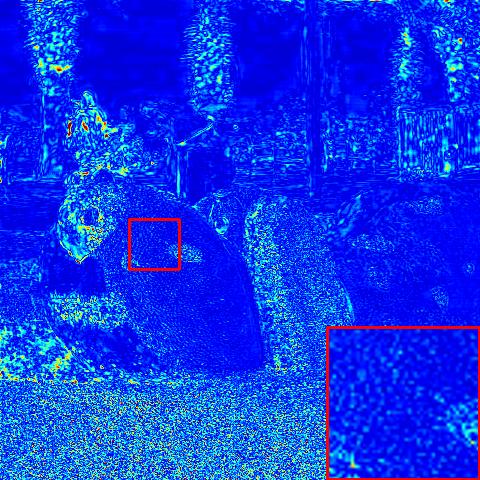}} \
			&{\includegraphics[width=2.42cm,height=2.42cm]{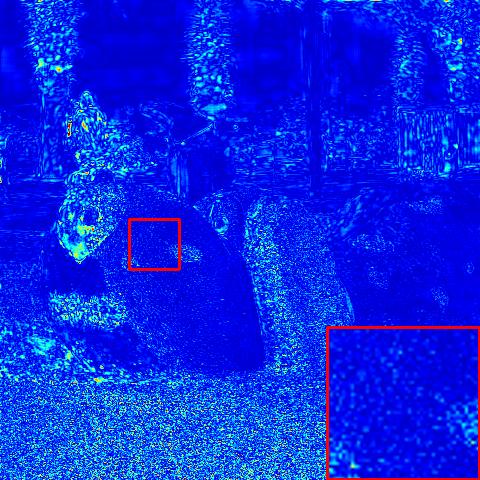}} \
			&{\includegraphics[width=2.42cm,height=2.42cm]{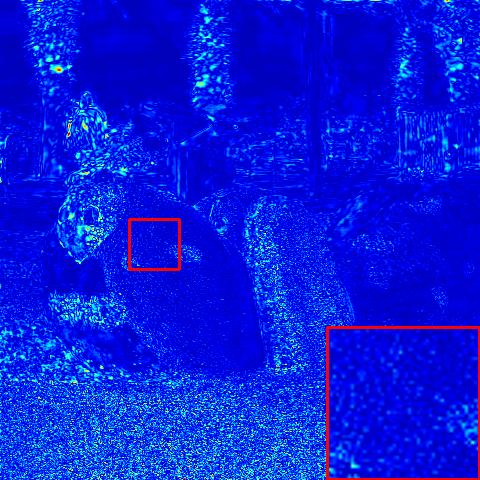}} \
			&{\includegraphics[width=2.42cm,height=2.42cm]{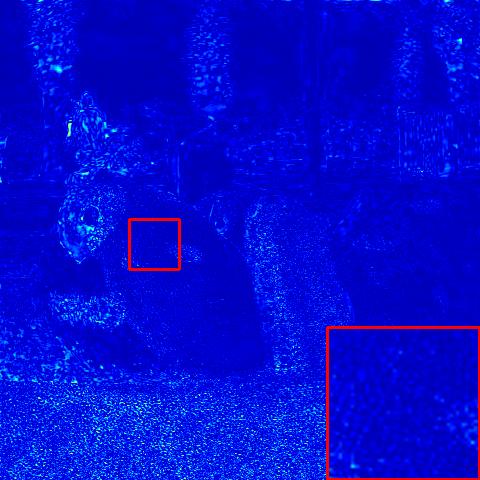}} \\	
		\end{tabular} 
	}
	\scalebox{2}
	{\includegraphics[width=0.51\textwidth]{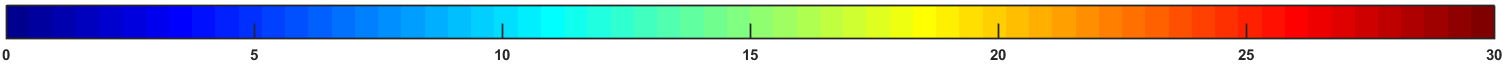}}
	\caption{Visual comparison on the ``BGU\_HS\_00263'' HSI (530nm) of NTIRE2018 ``Clean'', ``BGU\_HS\_00265'' HSI (550nm) of NTIRE2018 ``Real World'', ``ARAD\_HS\_00453'' HSI (570nm) of NTIRE2020 ``Clean'' and ``ARAD\_HS\_00457'' HSI (590nm) of  NTIRE2020 ``Real World'' datasets from top to bottom. The best view on the screen. 
	}
	\label{figure_results_heats}	
	%\vskip 0.05in
\end{figure*}

\subsection{Ablation Studies}
In this section, we conduct extensive ablation studies to thoroughly analyze the proposed HPRN. All experimental results are trained using the NTIRE2018 ``Clean'' training set and reported on the NTIRE2018 ``Clean'' validating data. The $baseline$ model is the backbone network consisting of $10$ MRBs, which only contain ordinary convolutional layers and trained by a single L1 loss constraint. For the TCRM, we explore its critical embedded location and the squeezed vector size. Correspondingly, we evaluate the effects of sharing embedding functions, the group size and the scales settings for the SSRM. Subsequently, the combined performances of SSRM, TCRM and SOPC are studied in detail.

\subsubsection{Explore the location of the TCRM} We investigate the influence of different positions for the developed TCRM and the results are listed in Table \ref{table_tcrm_pos}. The $\textit{1-pos}$, $\textit{2-pos}$ and $\textit{3-pos}$ denote before the first, second and third residual connection of the MRB, respectively. When our TCRM is embedded before the third residual summation of the MRB, it can produce better MRAE and SAM metrics than other two locations. The $\textit{multi-pos}$ indicates that the TCRM modules are added to the three residual positions. Although the $\textit{multi-pos}$ can further improve the performance, the parameters and computations are increased by three times over the single TCRM. Comprehensively, we choose the $\textit{3-pos}$ as our final scheme.

\subsubsection{Explore the squeezed vector size of the TCRM} As discussed above, the simple scalars are replaced with certain vectors as the squeezers of channel-wise relation to extract feature interdependences. Table \ref{table_tcrm_size} summarizes the impact of different squeezed vector sizes. ``Params'' and ``MACs'' denote the additional number of parameters and multiply-accumulate operations relative to the $baseline$ model. Results imply that $4\times4=16$ length vector can provide the more robust feature representations. Once the size of this vector further enlarge, for example $8\times8$, the model complexity also increase accordingly, which may lead to low-efficient Transformer-style feature interactions and reduce the accuracy of SSR.

\subsubsection{Effects of sharing embedding functions for the SSRM} Table \ref{table_ssrm_share} explores the effects of sharing weights of embedding functions for the SSRM. We can see that $\varphi(\cdot)=\psi(\cdot)$ not only achieves lower MRAE and SAM measurements but also saves the number of parameters. The main reason lies in that sharing embedding functions has a greater possibility to generate a more discriminative semantic-based relation matrix. Once the two embedding functions are different, the clustered-similar feature aggregation can be biased by their transformed projections. Therefore, we adopt the sharing embedding functions in the SSRM.

\subsubsection{Effects of the group size of the SSRM} The feature aggregation of SSRM is performed based on semantic-driven groups, hence the group size closely affects the effectiveness of SSRM. An appropriate $G$ value can assign category-consistent elements to each group as much as possible, so as to fulfill highly-effective correlation learning. To be specific, we set $G=\{16, 36, 64, 100, 144, 256\}$ and the experimental result is illustrated in Fig. \ref{figure_ssrm_group}. From the beginning, as the pixels involved in the correlation calculation increase, the error naturally continues to decrease. However, when $G$ is further enlarged, some newly introduced parts with little correlation would have a counterproductive effect, leading to a growing error. This shows that it is more crucial to choose informative features than to consider more. 

\subsubsection{Effects of the scales settings of the SSRM} Some errors on the clustering boundary may exist through running once SLIC algorithm with one scale (i.e. the number of categories). Moreover, different RGB signals may also be identified to inconsistent number of category attributes. Due to the two reasons, the multi-scales SLIC is employed parallelly, and the final multi-scales SSRM is weighted by each single scale one. Table \ref{table_ssrm_scale} reveals the scales settings of the SSRM. We try the four scales with various combinations to compare their effects, and adopt the scale settings of $8, 12, 16, 20$ based on the experimental results.

\begin{table*}
	\renewcommand\arraystretch{1.2}
	\centering
	\caption{The quantitative results of final test set of the NTIRE2018 ``Clean'' and ``Real World'' tracks. The best and second best results are \textbf{highlighted} and \underline{underlined}.}
	\setlength{\tabcolsep}{2mm}{
		\begin{tabular}{lcccccccccc}
			\toprule[1.4pt]
			\multirow{2}{*}{Model} & \multicolumn{5}{c}{$\text{Track 1:Clean}$} & \multicolumn{5}{c}{$\text{Track 2:Real World}$}  \\
			\cmidrule(r){2-6} \cmidrule(r){7-11}
			&$\text{MRAE($\downarrow$)}$&$\text{RMSE($\downarrow$)}$&$\text{SAM($\downarrow$)}$&$\text{PSNR($\uparrow$)}$&$\text{SSIM($\uparrow$)}$
			&$\text{MRAE($\downarrow$)}$&$\text{RMSE($\downarrow$)}$&$\text{SAM($\downarrow$)}$&$\text{PSNR($\uparrow$)}$&$\text{SSIM($\uparrow$)}$  \\
			\midrule[1.4pt]
			Sparse Coding\cite{arad2016sparse}                                        & 0.08094          & 59.41          & 5.021          & 37.23          & 0.97419          & {-------}             &{-------}           &{-------}          &{-------}           &{-------}             \\ %\hline
			Galliani\cite{galliani2017learned}& 0.05130 & 37.68 & 1.775 & 41.25 & 0.99606 & 0.06613    & 57.98 & 2.180 & 37.09 & 0.99155 \\ %\hline
			MSCNN\cite{yan2018accurate}& 0.03304 & 25.99 & 1.725 & 44.14 & 0.99748 & 0.04542    & 32.90 & 2.190 & 42.08 & 0.99405 \\ %\hline
			UNet\cite{stiebel2018reconstructing}  & 0.01552 & 15.39 & 1.144 & 49.10 & 0.99897 & 0.03151    & 24.15 & 1.730 & 44.88 & 0.99564 \\ %\hline
			FMNet\cite{zhang2020pixel}& 0.01404          & 13.22          & 1.010          & 50.35          &\underline{0.99923}   & 0.03053          & 23.43          & 1.660          & 45.15          & 0.99588\\
			HSCNN+\cite{shi2018hscnn+}                                      & \underline{0.01373}    & \underline{13.02}    &\underline{0.988}    &\underline{50.47}    & 0.99922          & 0.03042          & 23.60          & 1.667          & 45.10          & 0.99577\\
			HRNet\cite{zhao2020hierarchical}                                       & 0.01395          & 13.38          & 1.015          & 50.29          &\underline{0.99923}    &\underline{0.02915}    &\underline{22.84}    &\underline{1.580}    &\underline{45.43}    &\underline{0.99601}    \\
			Ours                                        & \textbf{0.01155} & \textbf{10.44} & \textbf{0.805} & \textbf{52.32} & \textbf{0.99943} & \textbf{0.02896} & \textbf{22.18} & \textbf{1.534} & \textbf{45.70} & \textbf{0.99603}\\
			\bottomrule[1.4pt]
	\end{tabular}}
	\label{table_test_2018}
\end{table*}

\begin{table*}
	\renewcommand\arraystretch{1.2}
	\centering
	\caption{The quantitative results of final test set of the NTIRE2020 ``Clean'' and ``Real World'' tracks. The best and second best results are \textbf{highlighted} and \underline{underlined}.}
	\setlength{\tabcolsep}{2mm}{
		\begin{tabular}{lcccccccccc}
			\toprule[1.4pt]
			\multirow{2}{*}{Model} & \multicolumn{5}{c}{$\text{Track 1:Clean}$} & \multicolumn{5}{c}{$\text{Track 2:Real World}$}  \\
			\cmidrule(r){2-6} \cmidrule(r){7-11}
			&$\text{MRAE($\downarrow$)}$&$\text{RMSE($\downarrow$)}$&$\text{SAM($\downarrow$)}$&$\text{PSNR($\uparrow$)}$&$\text{SSIM($\uparrow$)}$
			&$\text{MRAE($\downarrow$)}$&$\text{RMSE($\downarrow$)}$&$\text{SAM($\downarrow$)}$&$\text{PSNR($\uparrow$)}$&$\text{SSIM($\uparrow$)}$  \\
			\midrule[1.4pt]
			Sparse Coding\cite{arad2016sparse}& 0.07873 & 0.0331 & 5.572 & 31.12 & 0.96228& {-------}             &{-------}           &{-------}          &{-------}           &{-------}             \\ %\hline
			Galliani\cite{galliani2017learned}& 0.10720 & 0.0275 & 4.351 & 32.15 & 0.98237 & 0.11959    & 0.0352 & 4.761 & 30.19 & 0.96686 \\ %\hline
			MSCNN\cite{yan2018accurate}& 0.06835 & 0.0198 & 3.514 & 35.28 & 0.99085 & 0.09023    & 0.0214 & 4.271 & 34.25 & 0.97567 \\ %\hline
			UNet\cite{stiebel2018reconstructing}& 0.04389 & 0.0161 & 2.923 & 37.93 & 0.99444 & 0.07274    & 0.0193 & 3.725 & 35.42 & 0.98091 \\ %\hline
			HSCNN+\cite{shi2018hscnn+}& 0.04097 & 0.0154 & 2.776 & 37.90 & 0.99527 &\underline{0.07148}    & 0.0191 &\underline{3.668} & 35.45 & 0.98180\\
			FMNet\cite{zhang2020pixel}&\underline{0.04030} &\underline{0.0148} & \underline{2.688}&\underline{38.23} & 0.99595 & 0.07217    & 0.0190 & 3.744 &\underline{35.49}& 0.98215\\
			HRNet\cite{zhao2020hierarchical}& 0.04045 & 0.0154 & 2.729 & 38.16 &\underline{0.99601}& 0.07156    &\underline{0.0187}& 3.687 & 35.47 &\textbf{0.98284}\\
			Ours & \textbf{0.03784} & \textbf{0.0134} & \textbf{2.557} & \textbf{39.57} & \textbf{0.99654} & \textbf{0.07048}    & \textbf{0.0185} & \textbf{3.623} & \textbf{35.72} & \underline{0.98272}\\
			\bottomrule[1.4pt]
	\end{tabular}}
	\label{table_test_2020}
\end{table*}

\begin{table}[]
	\renewcommand\arraystretch{1.2}
	\centering
	\caption{Comparisons of computational parameters and running times on the NTIRE2018 ``Clean'' dataset.}
	\setlength{\tabcolsep}{0.6mm}{
		\begin{tabular}{@{}lccccccc@{}}
			\toprule[1.4pt]
			Method            & Galliani & MSCNN   & UNet    & FMNet   & HSCNN+  & HRNet   & Ours    \\ \midrule[1.4pt]
			Params(M)         & 2.124    & 26.687  & 1.044   & 6.948   & 19.551  & 31.705  & 14.893  \\
			Sizes(MB)         & 8.30    & 101  & 3.94   & 26.5   & 74.6  & 120  & 56.8  \\
			Times(ms) & 9.04     & 3.12    & 1.28    & 5.67    & 3.47    & 13.04   & 12.72   \\
			MRAE(↓)             & 0.05130  & 0.03304 & 0.01552 & 0.01404 & 0.01373 & 0.01395 & 0.01155 \\ \bottomrule[1.4pt]
	\end{tabular}}
	\label{table_time}
\end{table}

\begin{table}[]
	\renewcommand\arraystretch{1.2}
	\centering
	\caption{The classification performance on the UP dataset.}
	\setlength{\tabcolsep}{6.6mm}{
		\begin{tabular}{@{}lccc@{}}
			\toprule[1.4pt]
			Description            &OA($\uparrow$) & AA($\uparrow$)   &Kappa($\uparrow$)    \\ \midrule[1.4pt]
			Simulated RGB         & 86.07\%    & 81.12\%  & 80.87\%  \\
			Reconstructed HSI     & 88.07\%    & 83.14\%  & 83.52\%   \\
			Groundtruth HSI     & 88.89\%    & 85.91\%  & 84.91\%   \\
 \bottomrule[1.4pt]
	\end{tabular}}
	\label{table_remote}
\end{table}
 
\begin{figure*}[!tbp]
	\centering
	\subfloat[]{\includegraphics[width=0.23\textwidth]{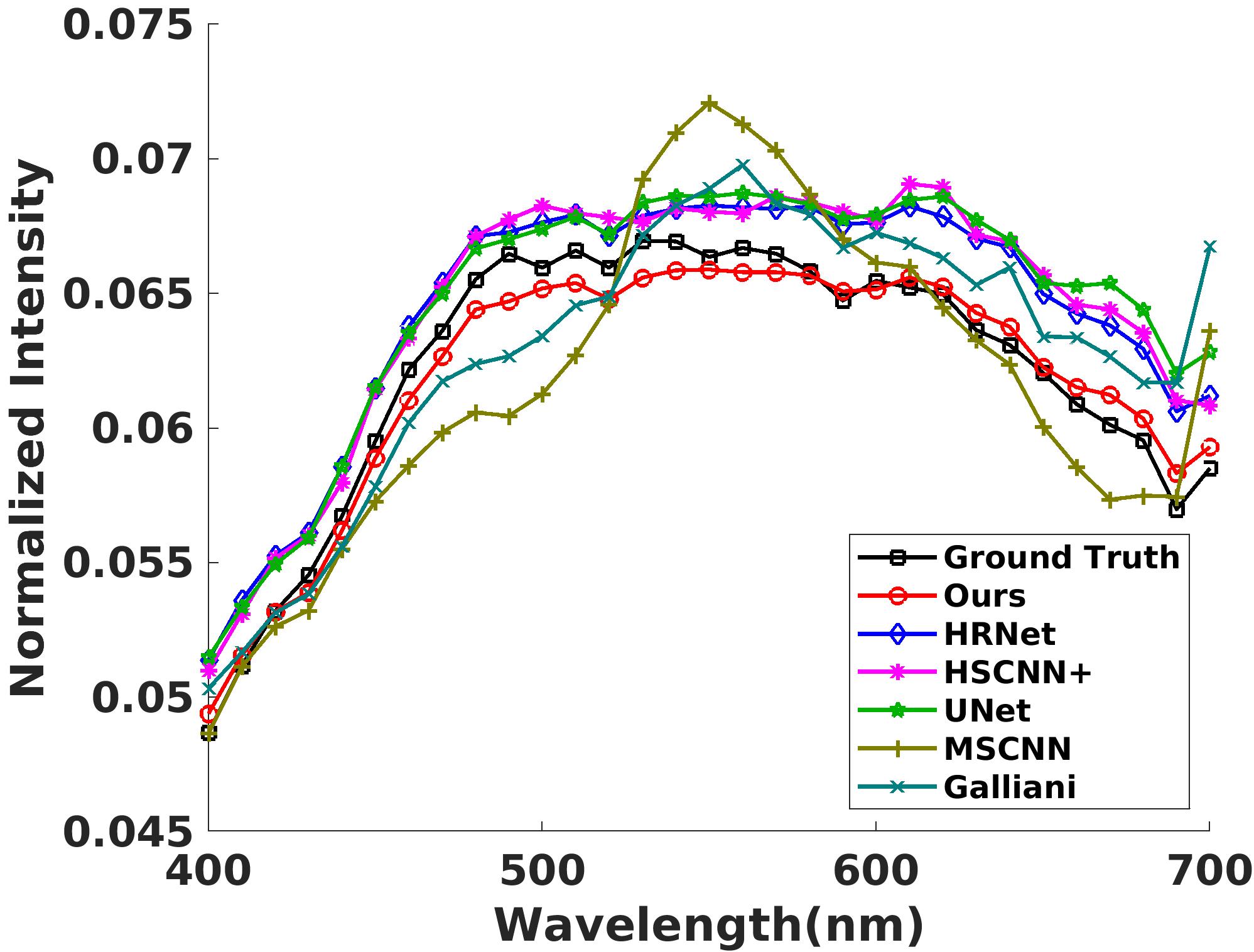}}\
	\subfloat[]{\includegraphics[width=0.23\textwidth]{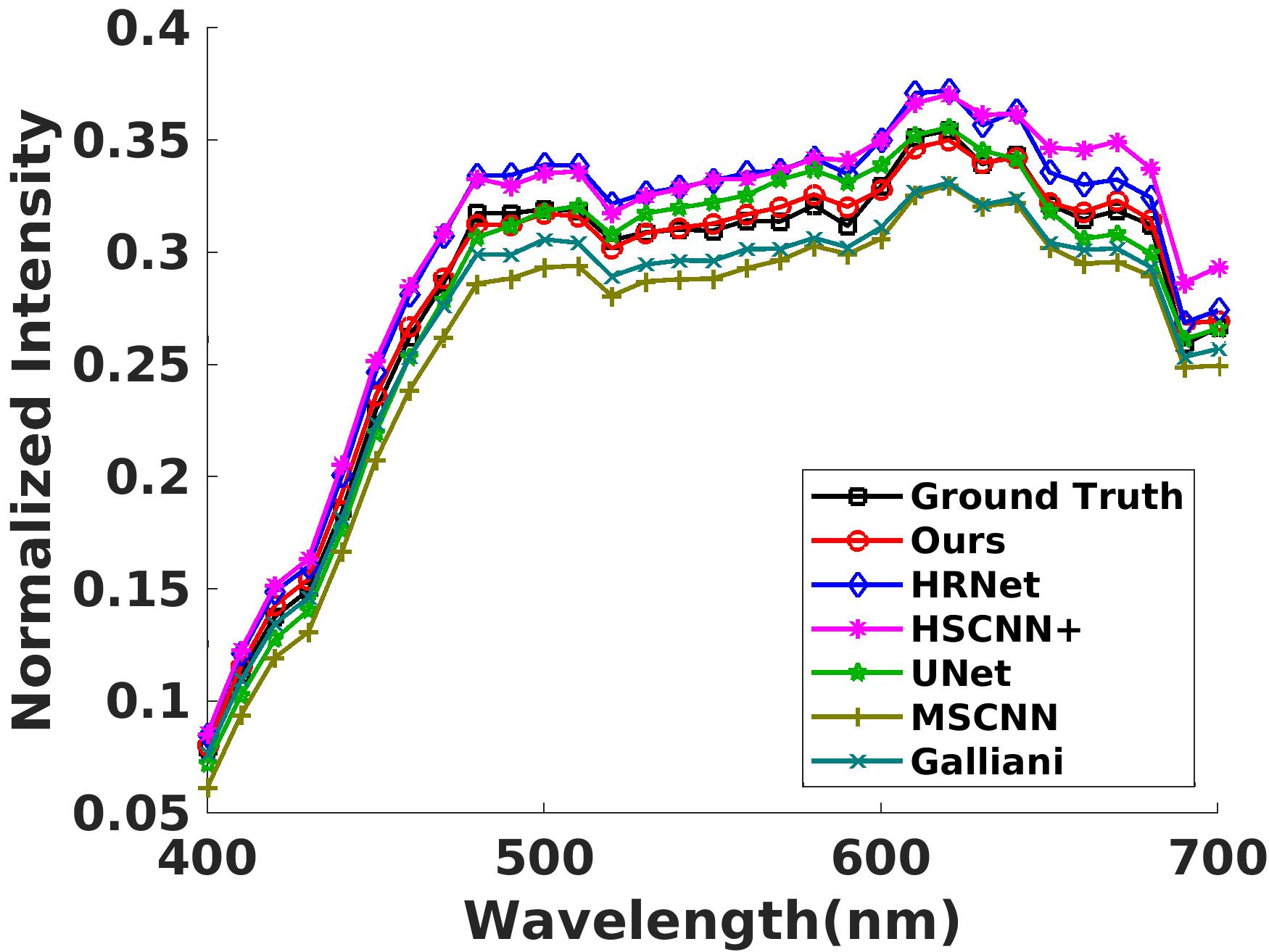}}\
	\subfloat[]{\includegraphics[width=0.23\textwidth]{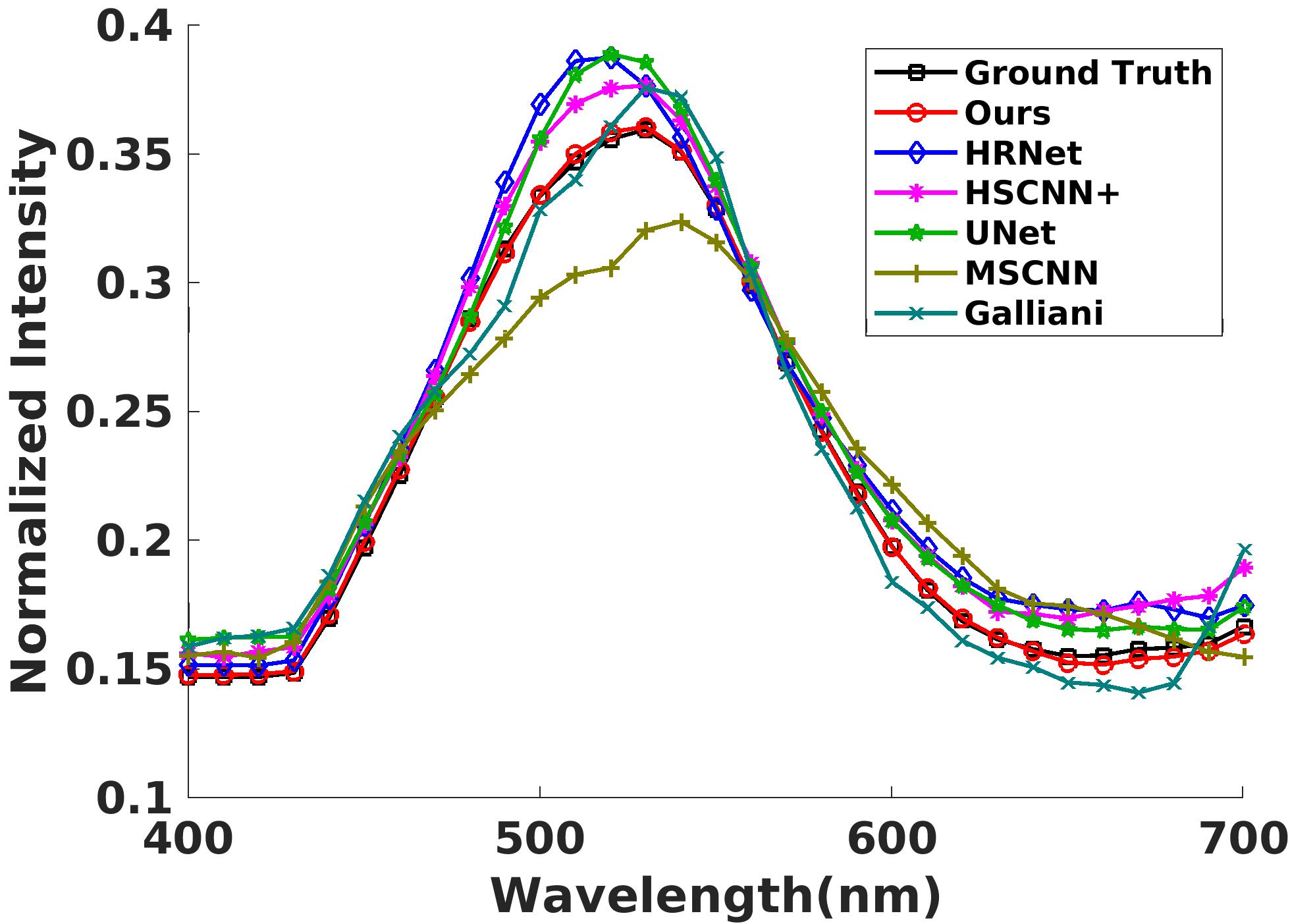}}
	\subfloat[]{\includegraphics[width=0.23\textwidth]{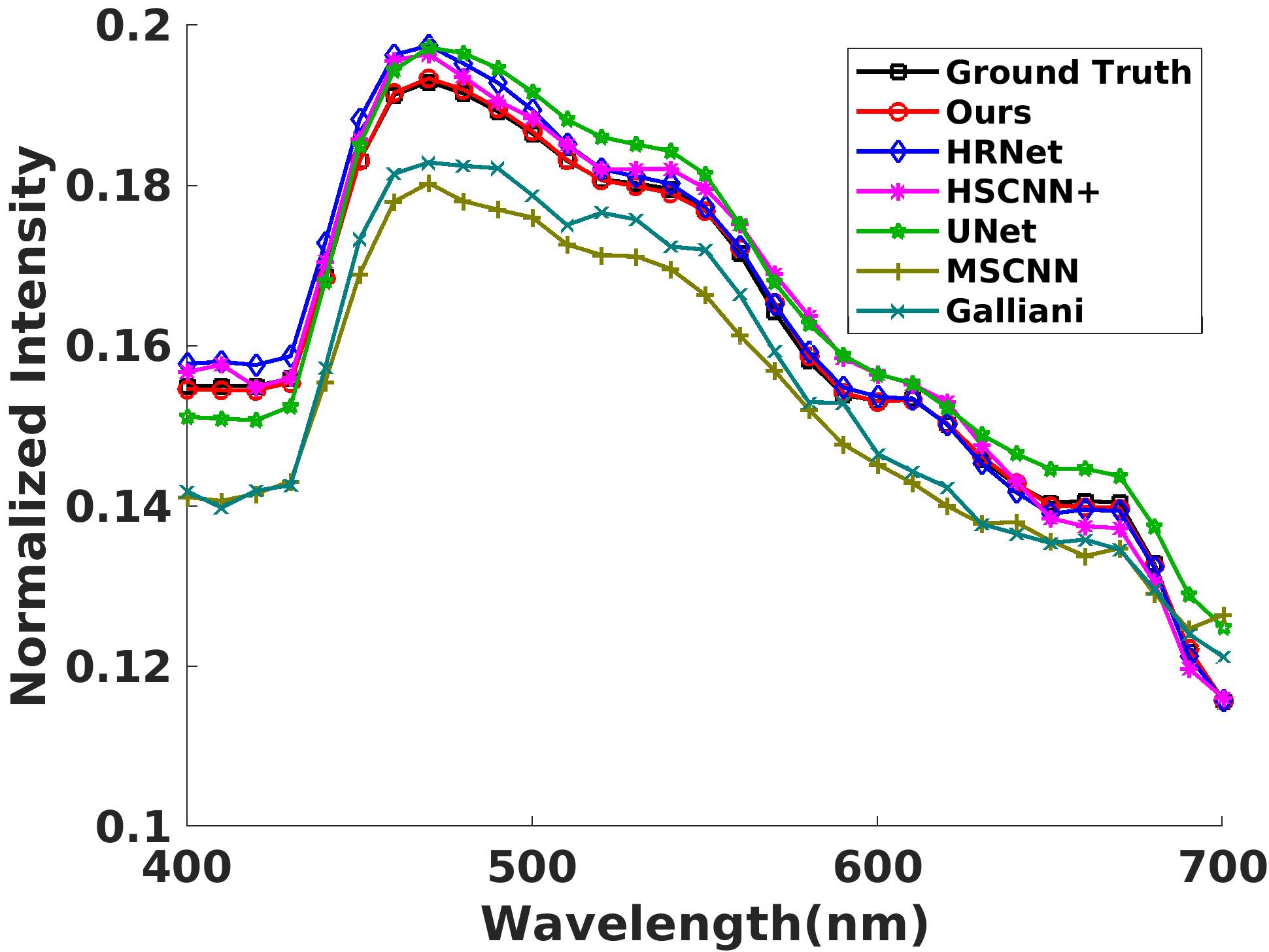}}\\
	\caption{Spectral response curves of selected several spatial points from the reconstructed HSIs. (a) and (b) are for the NTIRE2020 ``Clean'' and ``Real World'' tracks respectively. (c) and (d) are for the NTIRE2018 ``Clean'' and ``Real World'' track respectively.}
	\label{figure_results_curves}
\end{figure*}

\subsubsection{Effectiveness of different modules} The developed TCRM, well-designed SSRM and incorporated SOPC ensure that the proposed HPRN model can make the best of multi-source and abundant priors, including RGB semantic categories, deep feature-prior and HSIs band-wise correlations. To verify the effectiveness of these components, we evaluate the performance of HPRN with its variants in Table \ref{table_hprn}. Compared with the baseline MRAE and SAM scores, the results of $\textit{with TCRM}$, $\textit{with SSRM}$ and $\textit{with SOPC}$ successively prove the effectiveness of each individual section. On this basis, we combine the above three sections in pairs, and their results produce less error than using any single module. Finally, the presented HPRN model containing all parts achieves the best effect, which demonstrates that all modules are necessary for the proposed method to effectively alleviate the ill-posedness of this underconstrained SSR problem, and promote the accuracy of recovered HSIs.

\begin{figure*}[htbp]
	\centering
	\subfloat[]{\includegraphics[width=2.3cm,height=4.1cm]{pavia_intro/PaviaU_HSI_25_45_70.jpg}}\
	\subfloat[]{\includegraphics[width=2.3cm,height=4.1cm]{pavia_intro/Groundtruth_label.png}}\
	\subfloat[]{\includegraphics[width=2.3cm,height=4.1cm]{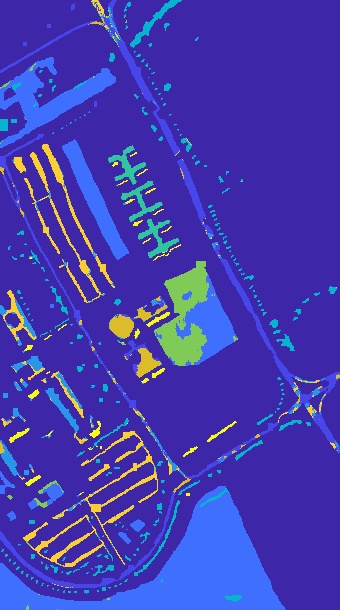}}\
	\subfloat[]{\includegraphics[width=2.3cm,height=4.1cm]{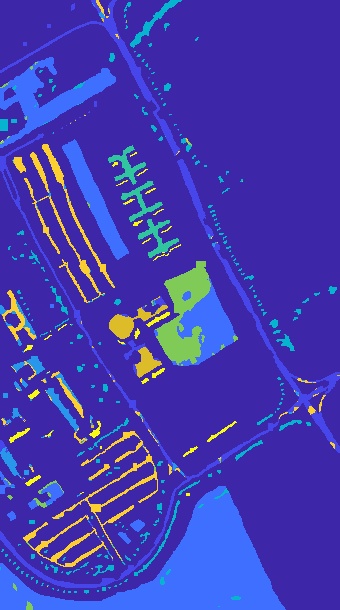}}\
	\subfloat[]{\includegraphics[width=2.3cm,height=4.1cm]{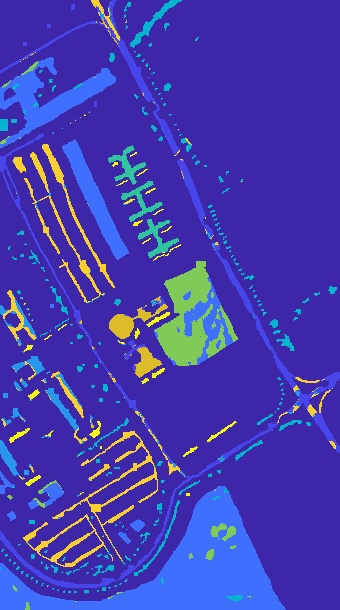}}\
	\subfloat[]{\includegraphics[width=3.3cm,height=4.1cm]{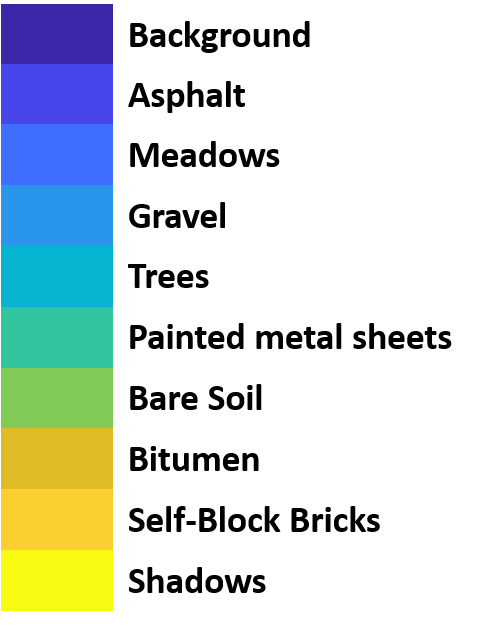}}\\
	\caption{Classification results on UP dataset. (a) False-color image. (b) Label. (c) Simulated RGB. (d) Reconstructed HSI. (e) Groundtruth HSI. (f) Categories.}
	\label{figure_pavia_result}
\end{figure*}

\subsection{Results on SSR Benchmark Datasets}
To evaluate the robustness and generalization of the proposed network, we compare with seven state-of-the-art methods, including Sparse Coding\cite{arad2016sparse}, Galliani\cite{galliani2017learned}, MSCNN\cite{yan2018accurate}, UNet\cite{stiebel2018reconstructing}, HSCNN+\cite{shi2018hscnn+}, FMNet\cite{zhang2020pixel} and HRNet\cite{zhao2020hierarchical}. For fair comparison, we run their released models on the training set of this paper. And the best model is chosen via the validation set, and further evaluated on the final testing set on four established benchmarks.

\subsubsection{Quantitative Results} The quantitative results of final test set of the NTIRE2018 and NTIRE2020 tracks are summarized in Table \ref{table_test_2018} and Table \ref{table_test_2020}, respectively. As we can see, our method consistently outperforms other approaches across the whole benchmarks in terms of five numerical metrics. From MRAE and RMSE assessments, our HPRN obtains the smallest values on all the datasets, which suggests that our estimating spectrum are the most accurate. It is worth noting that the proposed network improves SAM by 18.5\%, 2.9\%, 4.9\% and 1.2\% than the second best results on the NTIRE2018 ``Clean'', NTIRE2018 ``Real World'', NTIRE2020 ``Clean'' and NTIRE2020 ``Real World'' tracks. This indicates that our reconstructed HSIs contain better spectral continuity and authenticity. For structural measurement PSNR and SSIM, our method acquires larger improvement compared with these existing models. The reason may be that HPRN can adequately mine and utilize spatial contexts and semantic category priors of RGB images.

\subsubsection{Qualitative Evaluation} To further illustrate the superior performance of our network, Fig. \ref{figure_results_heats} displays certain examples of MRAE heat maps generated by our HPRN and other state-of-the-art methods. Visually, the bluer the displayed color, the lower the error of predicting HSIs. Overall, we can find that the presented HPRN produce the optimal visualizations, which are much close to the ground truth in various challenging scenarios. This visual performance is also in line with the numerical comparisons. Fig. \ref{figure_results_curves} describes the spectral response curves of selected several spatial points from the reconstructed HSIs. The change of the spectral curve is more representative of the essential attributes of the ground objects. From the qualitative evaluation, our recovered HSIs yield a spectral curve that is more consistent with the ground truth, which can also be reflected through the SAM criterion. To be specific, this observation infers that the SOPC plays an important role in maintaining the hyperspectral band-wise statistical correlations and spectral continuity.

\subsubsection{Model Efficiency Analyses} Table \ref{table_time} states the comparisons of computational parameters and running times of recent deep CNN-based SSR methods. The running time refers to the average inference time of multiple $512\times512$ images on an NVIDIA 2080ti GPU and an Intel Core i9-9900 CPU. Before the measurement, the machine will be warmed in advance. Due to the frequent index operations of SSRM, the testing time of the proposed HPRN reaches to 12.72ms. Among these networks, the parameters of Galliani and UNet are small, but the performance is not high. Compared with MSCNN, HSCNN+ and HRNet, our approach contains less parameters and better precision, which indicates that our method can balance the algorithm performance and model complexity.

\subsection{Results on Remote Sensing Dataset} To investigate the applicability of the reconstructed HSIs, we introduce the hyperspectral classification task to evaluate the SSR performance. Here we adopt the popular UP dataset to compare different classificational results, and the pixels are classified into nine classes through employing a classical CNN-based hyperspectral classification network \cite{sharma2016hyperspectral}. This model consists of three convolutional units and two fully-connected layers. And Table \ref{table_remote} shows the corresponding scores including overall accuracy (OA), average accuracy (AA), and Kappa coefficient (Kappa). We can observe that our reconstructed HSI achieves better OA, AA and Kappa than the simulated RGB image, and is closer to the classificational performance of using the groundtruth label, which is also reflected in Fig. \ref{figure_pavia_result}. This measurement demonstrates that the presented HPRN can effectively predict spectral information on the remote sensing dataset. Meanwhile, since the UP dataset contains 103 bands, which is more than the 31 bands of NTIRE2018 and NTIRE2020, this result also verifies that our algorithm has the ability to reconstruct more bands. To fulfill this, only the number of output channels of the last convolution layer needs to be modified according to the number of bands in the dataset.

\section{Conclusion}
In our paper, a novel holistic prior-embedded relation network (HPRN) is proposed for SSR. Multi-source and abundant priors including spatial contexts of RGB images, semantic categories of RGB signals, deep feature-prior and band-wise correlations of HSIs are incorporated into an end-to-end mapping function, which can effectively alleviate the ill-posedness of the SSR problem. To be specific, the backbone network is repeatedly stacked by several multi-residual relation blocks (MRBs), where the low-frequency context prior of RGB images is adequately utilized by the deep network. Through embedding semantic prior information of RGB inputs, a trainable semantic-driven spatial relation module (SSRM) is well-designed innovatively to perform the category-guided feature aggregation, and effectively achieve spectral optimization of the coarse estimation. Furthermore, a transformer-based channel relation module (TCRM) is investigated to leverage one local-range-averaging multiple-number vector to represent the channel of a deep feature. Coupled with Transformer-style feature interactions, the TCRM can obtain more discriminative learning power and make the RGB-to-HSI solution more robust. Finally, the second-order prior constraint (SOPC) is incorporated into the loss function to guide the HSI reconstruction to maintain the mathematical correlation and spectral consistency among hyperspectral bands. Extensive experimental results demonstrate that the presented HPRN can achieve superior performance on four SSR benchmarks under quantitative and perceptual comparisons. Also, the effectiveness of the estimating spectrum is verified by the classification results on the remote sensing dataset.

%	\appendices
%	\section{Proof of the First Zonklar Equation}
%	Appendix one text goes here.
%	
%	
%	\section{}
%	Appendix two text goes here.
	
%\section*{Acknowledgment}
	
	\ifCLASSOPTIONcaptionsoff
	\newpage
	\fi
	
	\bibliographystyle{IEEEtran}
	\bibliography{mybibfile}
	
\ifCLASSOPTIONcaptionsoff
\newpage
\fi
\vskip -0.5in
\begin{IEEEbiography}[{\includegraphics[width=1in,height=1.25in,clip,keepaspectratio]{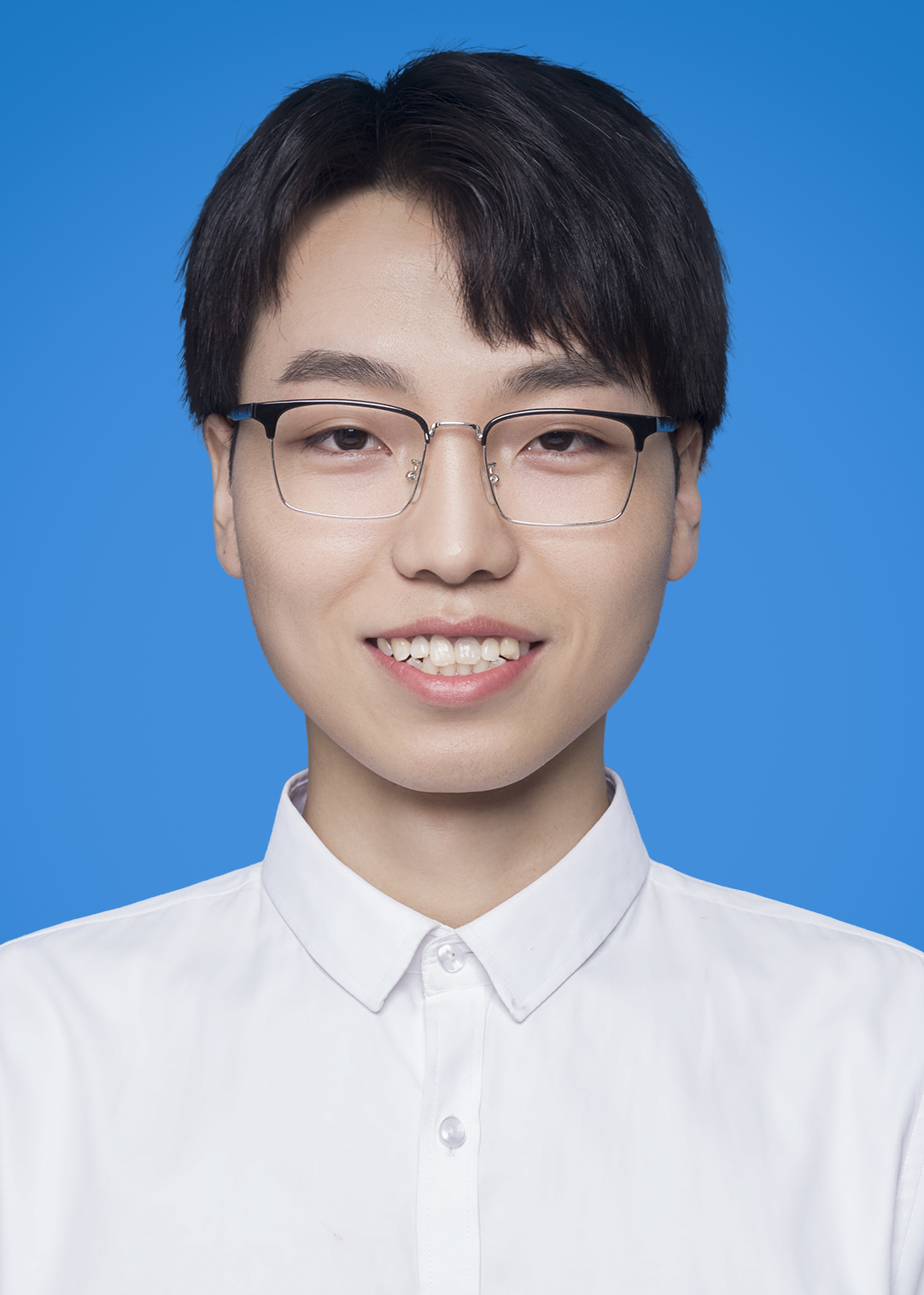}}]{Chaoxiong Wu}
	received the B.E. degree in telecommunications engineering from Xidian University, Xi’an, China, in 2018, where he is currently pursuing the Ph.D. degree with the State Key Laboratory of Integrated Service Network.
	His research interests include spectral super-resolution, semantic segmentation of remote sensing images and deep learning.
\end{IEEEbiography}
\vskip -0.5in
\begin{IEEEbiography}[{\includegraphics[width=1in,height=1.25in,clip,keepaspectratio]{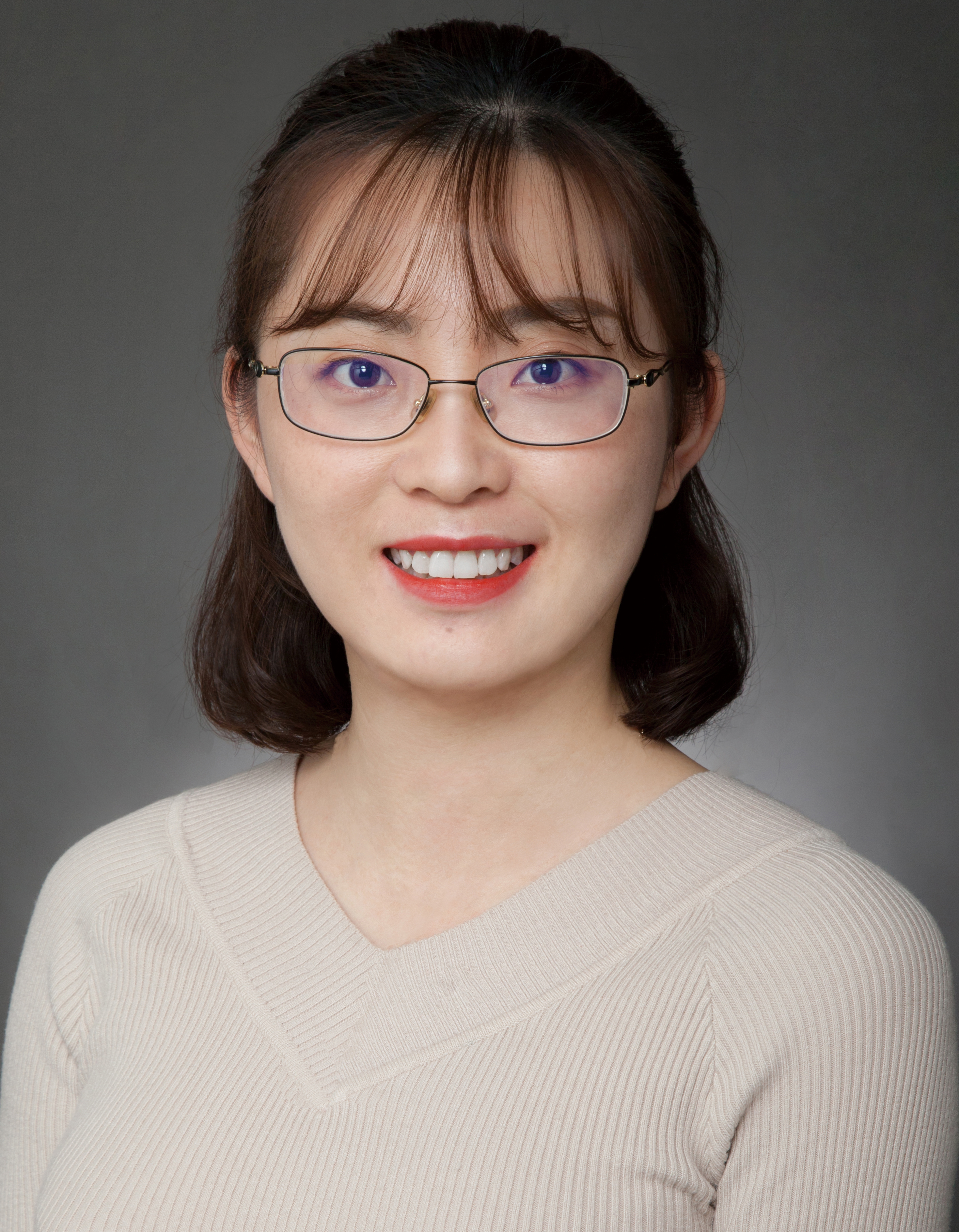}}]{Jiaojiao Li}
	(S'16-M'17) received the B.E. degree in computer science and technology, M. S. degree in software engineering and Ph.D. degree in communication and information systems from Xidian University in 2009, 2012 and 2016, respectively. She was an exchange Ph.D. Student of Mississippi State University supervised by Dr. Qian Du. She is currently an Associate Professor and Master supervisor with the school of Telecommunication, Xidian University, China. Her research interests include hyperspectral remote sensing image analysis and processing, pattern recognition.
\end{IEEEbiography}
\vskip -0.5in
\begin{IEEEbiography}[{\includegraphics[width=1in,height=1.25in,clip,keepaspectratio]{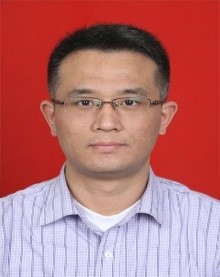}}]{Rui Song}
	received his Ph.D. degree in Signal and Information Processing from Xidian University, Xián, China in 2009. He is currently a Professor and Ph.D. advisor in the State Key Laboratory of Integrate Service Network, School of Tele-communications at Xidian University. His research interests include image and video coding algorithms and VLSI architecture design, intelligent image processing, and understanding and reconstruction of 3D scene.
\end{IEEEbiography}
\vskip -0.5in
\begin{IEEEbiography}[{\includegraphics[width=1in,height=1.25in,clip,keepaspectratio]{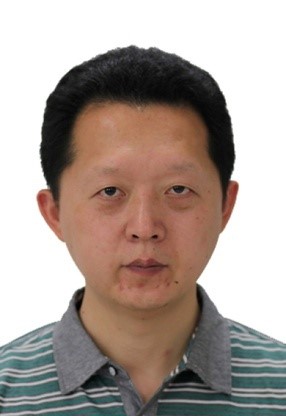}}]{Yunsong Li}
	received the M.S. degree in telecommunication and information systems and the Ph.D. degree in signal and information processing from Xidian University, China, in 1999 and 2002, respectively. He joined the school of telecommunications Engineering, Xidian University in 1999 where he is currently a Professor. Prof. Li is the director of the image coding and processing center at the State Key Laboratory of Integrated Service Networks. His research interests focus on image and video processing, hyperspectral image processing and high-performance computing.
\end{IEEEbiography}
\vskip -0.5in
%\vspace{-20mm}
\begin{IEEEbiography}[{\includegraphics[width=1in,height=1.25in,clip,keepaspectratio]{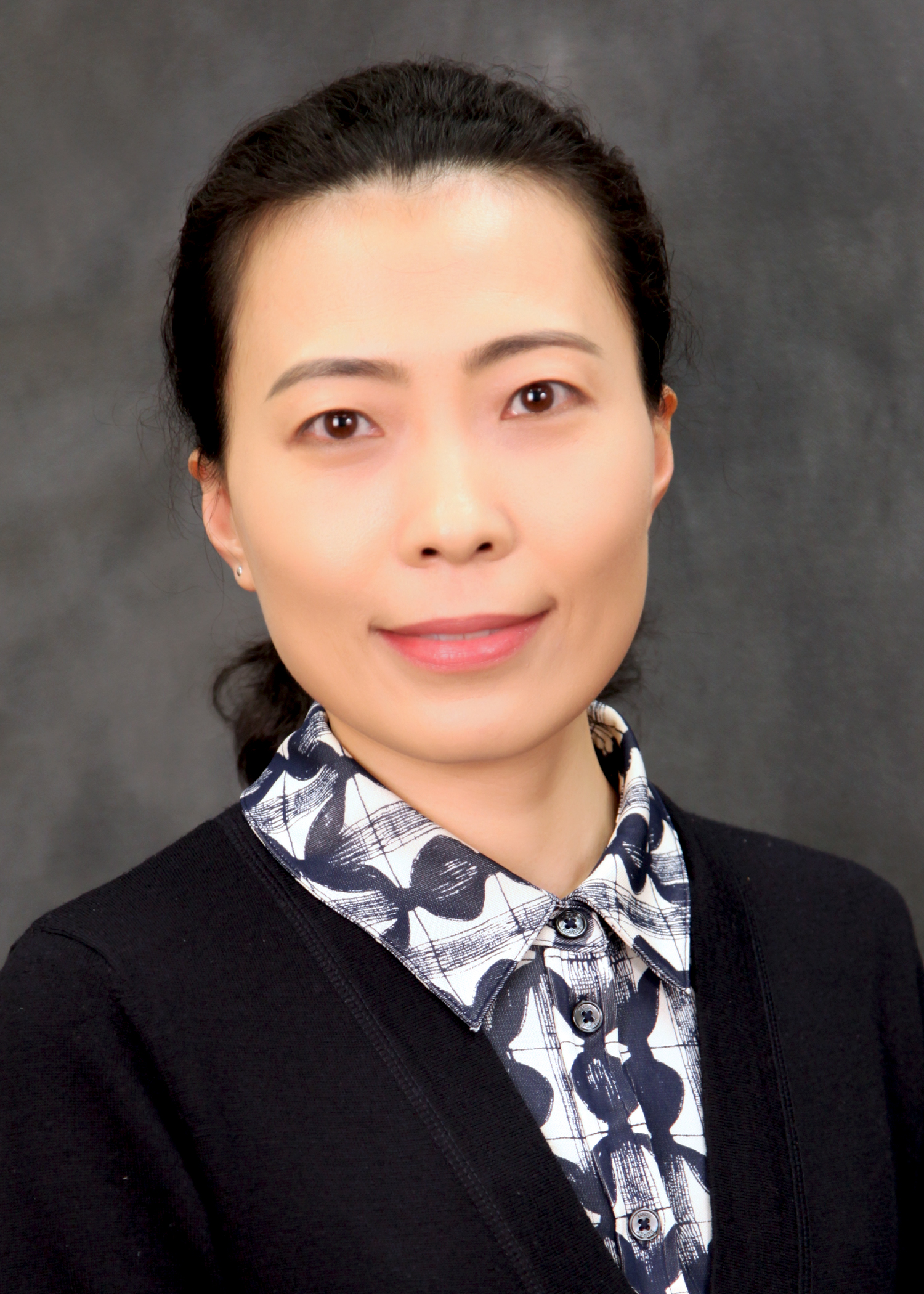}}]{Qian Du}
	(S’98–M’00–SM’05) received the Ph.D. degree in electrical engineering from the University of Maryland–Baltimore County, Baltimore, MD, USA, in 2000. She is currently a Bobby Shackouls Professor with the Department of Electrical and Computer Engineering, Mississippi State University, Starkville, MS, USA. Her research interests include hyperspectral remote sensing image analysis and applications, pattern classification, data compression, and neural networks.
	Dr. Du is a fellow of IEEE and SPIE–International Society for Optics and Photonics. She was a recipient of the 2010 Best Reviewer Award from the IEEE Geoscience and Remote Sensing Society (GRSS). She was as a Co-Chair for the Data Fusion Technical Committee of the IEEE GRSS from 2009 to 2013. She was the Chair with the Remote Sensing and Mapping Technical Committee of the International Association for Pattern Recognition from 2010 to 2014. She was the General Chair of the fourth IEEE GRSS Workshop on Hyperspectral Image and Signal Processing: Evolution in Remote Sensing held at Shanghai, China, in 2012. She served as an Associate Editor of the IEEE JOURNAL OF SELECTED TOPICS IN APPLIED EARTH OBSERVATIONS AND REMOTE SENSING (JSTARS), the Journal of Applied Remote Sensing, and the IEEE SIGNAL PROCESSING LETTERS. Since 2016, she has been the Editor-in-Chief of the IEEE JSTARS.
\end{IEEEbiography}

\end{document}